\documentclass[showpacs,preprintnumbers]{revtex4}
\usepackage{amsmath,amssymb,pstricks,wasysym,stmaryrd,enumerate,epsfig}
\usepackage[arrow,matrix,curve,ps]{xy}

\begin{document}
\title{Realisation of chiral symmetry in the domain model of QCD}
\author{Alex C. Kalloniatis\footnote{akalloni@physics.adelaide.edu.au} }

\address{
Special Research Centre for the Subatomic Structure of Matter,
University of Adelaide,
South Australia 5005, Australia}

\author{Sergei N. Nedelko \footnote{nedelko@thsun1.jinr.ru}}
\address{Institute of
Theoretical Physics~III of
Erlangen-Nuremberg University, Erlangen, Germany, and
Bogoliubov Laboratory of Theoretical Physics, JINR,
141980 Dubna, Russia}

\date{\today}
\preprint{ADP-03-141/T576}
\preprint{ FAU-TP3-03/11}

\begin{abstract}
The domain model for the QCD vacuum
has previously been developed and shown to exhibit confinement
of quarks and strong correlation of the local chirality of quark modes and 
duality of the background domain-like gluon field. 
Quark fluctuations satisfy a chirality violating boundary 
conditions parametrized by a random chiral angle $\alpha_j$ 
on the $j-th$ domain.
The free energy of an ensemble of $N\to\infty$ domains
depends on $\{ \alpha_j, j=1\dots N\}$ through the logarithm of
the quark determinant. Its
parity odd part is given by the axial anomaly. 
The anomaly contribution to the
free energy suppresses continuous axial $U(1)$ degeneracy in
the ground state, leaving only
a residual axial $Z(2)$ symmetry. This discrete symmetry 
and flavour $SU(N_f)_L\times SU(N_f)_R$ chiral symmetry
in turn are spontaneously broken
with a quark condensate arising due to the asymmetry
of the spectrum of Dirac operator.  
In order to illustrate the splitting between
the $\eta'$ from octet pseudoscalar mesons realised 
in the domain model, we estimate the masses of light pseudoscalar and 
vector mesons.

\end{abstract}
\maketitle

\section{INTRODUCTION}
A mechanism which simultaneously provides for
confinement of colour, spontaneously broken
chiral symmetry and a resolution of the $U_A(1)$
problem remains one of the open problems
in QCD today. Partial solutions \cite{vortex citations} based on specific
semi-classical or topologically stable configurations can go someway to
manifesting this triplet of phenomena but either
founder on generating all three or in allowing for
an effective model of the vacuum from
which hadron spectroscopy can be derived.
In any case, one expects that topological
objects of various dimensions - point-like, string-like, and sheet-like -
should contribute \cite{vanBaal} complete with
significant quantum fluctuations in a way that would be difficult
to describe via an interacting microscopic model. 
In this paper, we continue the 
exploration of the ``domain model'' for the vacuum, 
originally proposed  in \cite{NK2001}, as a scenario for
simultaneous appearance of all three phenomena:  confinement,
spontaneous chiral symmetry breaking via the appearance of
a quark condensate and a continuous
$SU(N_f)_L \times SU(N_f)_R$ degeneracy of the vacuum
for $N_f$ massless quarks, 
but without a $U_A(1)$ continuous
degeneracy of ground states that would be indicative of 
an unwanted Goldstone boson. 
The purpose of studying a model of this type is to 
identify the typical features of relevant nonperturbative gluonic
configurations. Such configurations would provide for as many 
gross features of nonperturbative QCD as possible. But this should
preserve simultaneously the well-studied short distance regime
and the model should be expressed in terms of quark-gluon 
degrees of freedom as well as in terms of colourless hadron bound states.

The model under consideration provides for confinement
of both static (area law) and dynamical  
(propagators are entire functions of momentum) 
quarks~\cite{NK2001}. It also displays specific chiral properties 
of quark eigenmodes.  Namely, as will be discussed in more detail below, 
the spectrum of the Dirac operator 
is asymmetric with respect to $\lambda\to-\lambda$  and
zero quark modes are absent, but the local chirality of all nonzero modes
at the centre of domains is correlated with the duality of 
the background field~\cite{NK2002}. This has been observed
on the lattice \cite{Xstudies} and is usually considered as an indication
of spontaneous breakdown of flavour chiral symmetry.
The purpose of this article is to study 
the details of chiral symmetry realization in the domain model. 
The nonzero quark condensate and axial anomaly are generated as a result of 
spectral asymmetry and definite mean chirality of
eigenmodes.
We compute the quark condensate, study the degeneracies of the 
minima of the free energy of the domain ensemble with respect to
chiral transformations and estimate the spectrum of pseudoscalar 
mesons.

The  model is defined by a partition function describing
an ensemble of hyperspherical domains, each characterised 
by a background covariantly constant self-dual or anti-self-dual
gluon field of random orientation. Summing over all orientations
and both self-dual and anti-self-dual fields 
guarantees Lorentz and CP invariance.
Quarks are confined as demonstrated in the original work \cite{NK2001}.
On the boundaries of each hypersphere, fermion fluctuations 
satisfy a chirality violating 
boundary condition
\begin{equation} 
i\!\not\!\eta(x) e^{i\alpha\gamma_5}\psi(x)=\psi(x)
\label{prelimbc}
\end{equation}
which is $2\pi$ periodic in the chiral angle $\alpha$.
Here $\eta_{\mu}$ is a unit radial vector at the boundary.  
Integrating over all such chiral angles guarantees
chiral invariance of the ensemble.  
As a consequence of Eq.~(\ref{prelimbc}) 
the spectrum of eigenvalues $\lambda$ of the Dirac operator
in a single domain is asymmetric under $\lambda\rightarrow -\lambda$.
Such asymmetries have been studied in other contexts,
for example by \cite{DGS98}.  
In the case of the domain model, the above boundary conditions are   
combined with the (anti-~)self-dual gluon field
which leads to a strong correlation between the
local chirality of quark modes at the centres of domains
with the duality of the background gluon field
\cite{NK2002}.
In this paper we study how these aspects contribute
to quark condensate formation and the pattern of chiral symmetry
breaking.

The vacua of the quantum problem associated with an 
ensemble of domains are the minima
of the free energy determined from the partition function.
The problem of the quark contributions to the free energy
requires calculation of the determinant of the Dirac operator 
in the presence of chirality violating boundary conditions.
For a choice of boundary 
condition with $\alpha\to -i \vartheta-\pi/2$ 
this problem has been tackled
in \cite{WD94} without a taking into account of the spectral asymmetry
where the parity odd part
of the logarithm of the determinant was identified as 
$\ln {\rm det} (i\!\not\!D) \sim   2q\vartheta$ 
with  $q$ the topological charge (not necessarily integer) 
of the underlying gluon field, namely the axial anomaly.

Our first goal in this paper is to address the analogous problem
for the specific gluon field relevant to
the domain model taking into account the asymmetry of the spectrum. 
For the parity odd part we obtain
\begin{equation}
\ln {\rm det} (i\!\not\!D) \sim 2 i q (\alpha \ {\rm mod} \  \pi).
\label{prelimanomaly}
\end{equation}
This result is consistent with \cite{WD94} up to a contribution coming from 
the asymmetry spectral function. However, we obtain an additional
parity even part which also turns out to be $\alpha$-dependent. 
We consider this to be more an artifact of the incompleteness
of our calculation than an established property of
the determinant. 

In the partition function all possible sets of chiral angles 
$\left\{\alpha_1,\dots,\alpha_N \right\}$
are summed ensuring the chiral invariance of the ensemble. 
Summation over all degrees of freedom besides chiral angles
defines the free energy as a function of these chiral angles.
In the limit $N\to \infty$  the minima of the free energy density 
in $\left\{\alpha_1,\dots,\alpha_N \right\}$ determine the preferred
chiral angles. More specifically,
when self-dual and anti-self-dual configurations are summed,
the anomaly Eq.~(\ref{prelimanomaly}) leads to a contribution
to the free energy of the form $-\ln\cos(2q\arctan(\tan\alpha))$ which
vanishes when $\alpha=0,\pi$.  The minima are degenerate with respect to 
discrete $Z_2$ chiral transformations. Each of these minima are 
characterised by a quark condensate of opposite sign, 
which arises due to the the spectral asymmetry.
An infinitesimally small
quark mass removes the degeneracy between the two discrete minima,
and a nonzero quark condensate is generated with value
\begin{equation*}
\langle \bar \psi(x) \psi(x)\rangle  = -(237.8 \ {\rm MeV})^3
\end{equation*}
with no additional modifications of the two model parameters
after fixing in the gluonic sector of the theory.
This gives a model with the chiral $Z_2$  discrete subgroup of $U_A(1)$ 
being spontaneously broken, and not the continuous $U_A(1)$ itself.
In the absence of the mass term the ensemble average 
of $\bar\psi \psi$ correctly vanishes. 
 A similar argument based on minimisation of the free energy 
and thereby a relaxation of the effective $\theta$-parameter of QCD to zero 
is discussed in detail in~\cite{Mink} in the context of the
strong CP-problem.

Moreover, the form of Eq.(\ref{prelimanomaly})
means that the free energy does not depend on flavour nonsinglet 
chiral angles when more than one massless quark flavours are introduced.  
This allows for the correct degeneracy of vacua with respect to continuous 
$SU(N_f)_L\times SU(N_f)_R$
chiral transformations. This vacuum structure implies 
the existence of Goldstone bosons in the flavour nonsinglet 
pseudoscalar channel but not in the singlet channel. 
To unveil
more explicitly the  singlet-octet splitting  we analyse 
the structure of pseudoscalar correlation functions in
the context of the domain model and estimate the masses of light pseudoscalar 
and vector mesons. The qualitative conclusion of this analysis is that 
the area law (confinement of static quarks) and the singlet-octet 
splitting in the model have the same origin: the finite range 
correlations of the background gluon field.

In the next section we briefly review the model, followed by
a summary of the properties of the spectrum of the Dirac 
operator in the domain-like gluon field. We then discuss in
detail the calculation of the logarithm of the
quark determinant for one massless quark flavour, including the
role of spectral asymmetry in domains in giving the anomaly
for the parity odd part. This is followed by an analysis of
the symmetries of the ground state of the domain ensemble
and the computation of the condensate. In section V we generalise
the result to $N_f$ massless flavours in order to verify the
spontaneous breaking of $SU(N_f)_L\times SU(N_f)_R$ in the
ensemble. 
The last section is devoted to calculation of meson masses.
Details of calculations
are relegated to the appendix.

\section{THE DOMAIN MODEL}
For motivation and a detailed description of the model we refer the 
reader to \cite{NK2001}.
The essential definition of the model
is given in terms of the following partition function for
$N\to\infty$ domains of radius $R$
\begin{eqnarray}
{\cal Z} =  {\cal N}\lim_{V,N\to\infty}
\prod\limits_{i=1}^N
\int\limits_{\Sigma}d\sigma_i
\int_{{\cal F}_\psi^i}{\cal D}\psi^{(i)} {\cal D}\bar \psi^{(i)}
\int_{{\cal F}^i_Q} {\cal D}Q^i 
\delta[D(\breve{\cal B}^{(i)})Q^{(i)}]
\Delta_{\rm FP}[\breve{\cal B}^{(i)},Q^{(i)}]
e^{- S_{V_i}^{\rm QCD}
\left[Q^{(i)}+{\cal B}^{(i)}
,\psi^{(i)},\bar\psi^{(i)}
\right]}
\label{partf}
\end{eqnarray}
where the functional spaces of integration
${\cal F}^i_Q$ 
and ${\cal F}^i_\psi$  are specified by the boundary conditions  
$(x-z_i)^2=R^2$
\begin{eqnarray}
\label{bcs}
&&\breve n_i Q^{(i)}(x)=0, 
\\
&&i\!\not\!\eta_i(x) e^{i\alpha_i\gamma_5}\psi^{(i)}(x)=\psi^{(i)}(x),
\label{quarkbc} \\
&&\bar \psi^{(i)} e^{i\alpha_i\gamma_5} i\!\not\!\eta_i(x)=-\bar\psi^{(i)}(x).
\label{adjquarkbc} 
\end{eqnarray}
Here $\breve n_i= n_i^a t^a$ with the 
generators $t^a$  of $SU_{\bf c}(3)$ in the adjoint representation
and the $\alpha_i$ are chiral angles 
associated with the boundary condition
Eq.(\ref{quarkbc}) with different values randomly assigned to domains. 
We shall discuss this constraint in detail in later sections.
The thermodynamic limit assumes $V,N\to\infty$ but 
with the density $v^{-1}=N/V$ taken fixed and finite. The
partition function is formulated in a background field gauge
with respect to the domain mean field, which is approximated 
inside and on the boundaries of the domains by
a covariantly constant (anti-)self-dual gluon field with the 
field-strength tensor of the form
\begin{eqnarray*}
F^{a}_{\mu\nu}(x)
&=&
\sum_{j=1}^N n^{(j)a}B^{(j)}_{\mu\nu}\vartheta(1-(x-z_j)^2/R^2), \nonumber \\
B^{(j)}_{\mu\nu}B^{(j)}_{\mu\rho}&=&B^2\delta_{\nu\rho}.
\end{eqnarray*}
Here  $z_j^{\mu}$ are the positions of the centres of domains in 
Euclidean space.

The measure of integration over parameters characterising domains is 
\begin{eqnarray}
\label{measure}
\int\limits_{\Sigma}d\sigma_i\dots =  \frac{1}{48\pi^2}
\int_V\frac{d^4z_i}{V}
\int_{0}^{2\pi}d\alpha_i
\int\limits_0^{2\pi}d\varphi_i\int_0^\pi d\vartheta_i\sin\theta_i
\int_0^{2\pi} d\xi_i
\sum\limits_{l=0,1,2}^{3,4,5}
\delta(\xi_i-\frac{(2l+1)\pi}{6})
\int_0^\pi d\omega_i\sum\limits_{k=0,1}\delta(\omega_i-\pi k)
\dots ,
\end{eqnarray}
where $(\theta_i,\varphi_i)$ are the spherical angles of the 
chromomagnetic field, $\omega_i$ is the angle between chromoelectric and 
chromomagnetic fields and $\xi_i$ is an angle parametrising the colour 
orientation. 

This partition function describes a statistical system
of the domain-like structures, of density $v^{-1}$ where
the volume of a domain is $v=\pi^2R^4/2$. Each domain is
characterised by a set of internal parameters and
whose internal dynamics are represented by fluctuation fields.
Most of the symmetries of the QCD Lagrangian are respected,
since the statistical ensemble is invariant under space-time and 
colour gauge transformations.
For the same reason, if the quarks are massless then the
chiral invariance is respected.
The model involves only
two free parameters: the mean field strength $B$ and the
mean domain radius $R$.
These dimensionful parameters break the scale invariance
present originally in the QCD Lagrangian.
In principle, they should be
related to the trace anomaly of the energy-momentum tensor \cite{Nie77, Mink81}
and, eventually, to the fundamental scale $\Lambda_{QCD}$.
Knowledge of the full quantum effective action of QCD would be 
required for establishing a relation of this kind.

A straightforward application of Eq.~(\ref{partf}) 
to the vacuum expectation value of a product of $n$
field strength tensors, each of the form
$$
 B^{a}_{\mu\nu}(x)
=\sum_j^N n^{(j)a}B^{(j)}_{\mu\nu}\theta(1-(x-z_j)^2/R^2),
 $$
gives for the connected $n$-point
correlation function
\begin{eqnarray*}
\langle B^{a_1}_{\mu_1\nu_1}(x_1)\dots B^{a_n}_{\mu_n\nu_n}(x_n) \rangle
& =&
\lim_{V,N\to\infty}\sum_j^N\int_V\frac{dz_j}{V}\int d\sigma_j
n^{(j)a_1}\dots n^{(j)a_n}B^{(j)}_{\mu_1\nu_1}\dots B^{(j)}_{\mu_n\nu_n}
\nonumber\\
&\times&\theta(1-(x_1-z_j)^2/R^2)\dots
\theta(1-(x_n-z_j)^2/R^2)
\nonumber\\
&=& B^{n} t^{a_1\dots a_n}_{\mu_1\nu_1,\dots,\mu_n\nu_n}
\Xi_n(x_1,\dots,x_n),
\end{eqnarray*}
where the tensor $t$ is given by the integral
$$
t^{a_1\dots a_n}_{\mu_1\nu_1,\dots,\mu_n\nu_n}=
\int d\sigma_j
n^{(j)a_1}\dots n^{(j)a_n}B^{(j)}_{\mu_1\nu_1}\dots B^{(j)}_{\mu_n\nu_n},
$$ 
and can be calculated explicitly using the measure, Eq.~(\ref{measure}). 
This tensor vanishes for odd $n$. In particular, the integral over 
spatial directions is defined by the generating formula
\begin{eqnarray*}
\frac{1}{4\pi}
\int\limits_0^{2\pi}d\varphi_j\int_0^\pi d\theta_j\sin\theta_j
e^{B^{(j)}_{\mu\nu}J_{\mu\nu}}=
\frac{\sin\sqrt{2B^2[J_{\mu\nu}J_{\mu\nu}\pm 
\tilde J_{\mu\nu}J_{\mu\nu}]}}{\sqrt{2B^2[J_{\mu\nu}J_{\mu\nu}\pm 
\tilde J_{\mu\nu}J_{\mu\nu}]}}.
\end{eqnarray*}
The translation-invariant function
\begin{eqnarray}
\label{corfunctn}
\Xi_n(x_1,\dots,x_n)=\frac{1}{v}\int d^4z
\theta(1-(x_1-z)^2/R^2)\dots
\theta(1-(x_n-z)^2/R^2)
\end{eqnarray}
can be seen as the volume of the region of overlap of $n$ 
hyperspheres of radius $R$ and centres ($x_1,\dots,x_n$),
normalised to the volume of a single hypersphere
$v=\pi^2R^4/2$,
$$
\Xi_n=1, \ {\rm for} \ x_1=\dots=x_n.
$$ 
It is obvious from this geometrical interpretation 
that $\Xi_n$ is a continuous function and 
vanishes if the distance between any two points
$|x_i-x_j|\ge 2R$; correlations in the background field have finite range
$2R$. The Fourier transform of $\Xi_n$ is then an entire analytical 
function and thus correlations do not have a particle interpretation.
It should be stressed that the statistical ensemble 
of background fields is not Gaussian since all connected correlators
are independent from each other and cannot be reduced 
to the two-point correlations. 

Within this framework the gluon condensate to lowest order in fluctuations
is $4B^2$, the absolute value of the topological charge per 
domain reads $q=B^2R^4/16$ and the topological  susceptibility turns out to 
be $\chi=B^4R^4/128\pi^2$.
An area law is obtained for static quarks. 
Computation of the Wilson loop for a circular contour  of
a large  radius $L\gg R$ gives a string tension $\sigma = B f(\pi B R^2)$
where $f$ is given for colour $SU(2)$ and $SU(3)$ in \cite{NK2001}.
The area law emerges due to the finite range of background field correlators 
Eq.~(\ref{corfunct}).
On the other hand, the model cannot account for such a subtle feature as 
Casimir scaling:
the adjoint Wilson loop naturally shows perimeter law, but trivially  
because of the abelian character 
of the domain mean field.
 
Estimations of the values of these quantities are known from 
lattice calculation or phenomenological approaches and can be used to fit  
$B$ and $R$.
As described in \cite{NK2001} these parameters are fixed to
be
$\sqrt{B} = 947 {\rm {MeV}}, R=(760 {\rm{MeV}})^{-1} = 0.26 {\rm {fm}}$ 
with the average absolute value of topological charge per domain
turning out to be $q\approx 0.15$ and the density of domains 
$v^{-1}=42{\rm fm}^{-4}$. The topological  susceptibility is then
 $\chi = (197 {\rm MeV})^4$, comparable to the 
Witten-Veneziano value \cite{largeNc}. This fixing of the parameters
of the model remains unchanged in this investigation of the quark
sector.  The quark condensate
at the origin of a domain where angular dependence drops out was estimated 
in paper \cite{NK2001} with 
a result of $-(228 {\rm{MeV}})^3$. 

\section{DIRAC OPERATOR AND SPECTRUM}
 
The eigenvalue problem 
\begin{eqnarray*}
\!\not\!D\psi(x)&=&\lambda \psi(x),
\nonumber\\
i\!\not\!\eta(x) e^{i\alpha\gamma_5}\psi(x)&=&\psi(x), \ x^2=R^2
\end{eqnarray*}
was studied in \cite{NK2002}.
Dirac matrices are in an anti-hermitean representation.
For $\alpha$ assumed to be real a bi-orthogonal basis has to 
be constructed.
Solutions can be labelled
via the Casimirs and eigenvalues 
\begin{eqnarray*}
{\bf K}_1^2 & = & {\bf K}_2^2 \rightarrow 
            \frac{k}{2}(\frac{k}{2} + 1), \  k=0,1,\dots,\infty \\
K^z_{1,2} & \rightarrow & m_{1,2}, 
\nonumber \\
 m_{1,2}&=&-k/2, -k/2+1,\dots,k/2-1,k/2,
\end{eqnarray*}
corresponding to the angular momentum operators
\begin{eqnarray*}
{\bf K}_{1,2} = \frac{1}{2} ({\bf L} \pm {\bf M})
\end{eqnarray*}
with ${\bf L}$ the usual three-dimensional angular momentum
operator and ${\bf M}$ the Euclidean version of the boost operator.
The solutions for the self-dual background field are then 
\begin{eqnarray}
\psi^{-\kappa}_{km_1}=i\!\not\!\eta\chi^{-\kappa}_{km_1}
+\varphi^{-\kappa}_{km_1},
\label{psikappa}
\end{eqnarray}
where $\chi$ and $\varphi$ must both have negative chirality in the
self-dual field and $\kappa$ is related to the polarisation of the field 
defined via the projector
\begin{eqnarray}
O_{\kappa} = N_+ \Sigma_{\kappa} + N_- \Sigma_{-\kappa}
\end{eqnarray}
with 
\begin{eqnarray*}
N_{\pm} = \frac{1}{2}(1\pm \hat n/|\hat n|), \
\Sigma_\pm = \frac{1}{2}(1\pm{ \bf\Sigma  B}/B)
\end{eqnarray*}
being respectively separate projectors for colour and spin polarizations.
 Significantly, the negative chirality for $\chi$ and $\varphi$ is
the only choice for which the boundary condition Eq.(\ref{quarkbc})
can be implemented for the self-dual background. 
The explicit form of the spinors $\chi$ and
$\varphi$ can be found in \cite{NK2002}, where it is demonstrated
that the eigenspinor Eq.~(\ref{psikappa}) has definite chirality at the centre
of domain correlated with the duality of the gluon field.
The boundary condition reduces to
\begin{eqnarray}
\label{bc-main}
\chi=- e^{ \mp i\alpha}\varphi ,  \  \bar\chi= \bar\varphi e^{ \mp i\alpha},
\ x^2=R^2,
\end{eqnarray}
where upper (lower) signs correspond to  $\varphi$ and $\chi$
with chirality $\mp1$, which, using the solutions, amounts to
equations for the two possible polarisations, for $\Lambda^{-+}_{k}$:
\begin{eqnarray}
\label{L-+}
e^{-i\alpha}M\left(k+2-\Lambda^2,k+2,z_0\right)
-\frac{\sqrt{z_0}}{i\Lambda}
\left[ M\left(k+2-\Lambda^2,k+2,z_0\right)
-\frac{k+2-\Lambda^2}{k+2}M\left(k+3-\Lambda^2,k+3,z_0\right)
\right] =0,
\end{eqnarray}
and for $\Lambda^{--}_{k}$:
\begin{eqnarray}
\label{L--}
e^{-i\alpha}M\left(-\Lambda^2,k+2,z_0\right)
+\frac{i\Lambda \sqrt{z_0}}{k+2}
M\left(1-\Lambda^2,k+3,z_0\right)
=0
\end{eqnarray}
where $z_0 = \hat BR^2/2$ and $\Lambda=\lambda/\sqrt{2\hat B}$.
 For the present work
Eqs.(\ref{L-+},\ref{L--}) are the starting point, from
which we see by inspection that 
a discrete spectrum of complex eigenvalues emerges
for which there is
no symmetry of the form $\lambda\rightarrow -\lambda$.
For given
chirality and polarisation and angular momentum $k$,
an infinite set of discrete $\Lambda$ are obtained labelled
by a ``principal quantum number'' $n$.

\section{QUARK DETERMINANT AND FREE ENERGY FOR A SINGLE DOMAIN}

\subsection{Massless case} 
We consider the one-loop contribution of the quarks to the free energy 
density $F(B,R|\alpha)$
of a single (anti-)self-dual domain of  volume $v=\pi^2 R^4/2$ 
\begin{eqnarray}
\exp\left\{-v F(B,R|\alpha)\right\}
&=&{\rm det}_\alpha\left(\frac{i\!\not \!D}{i\!\not\! \partial}\right)
\nonumber  \\
&=&\prod_{\kappa,k,n,m_1}\left(
\frac{\lambda^\kappa_{kn}(B)}{\lambda^\kappa_{kn}(0) }\right) \nonumber \\
&=&\exp\left\{-\zeta'(s)\right\}_{s=0}.
\label{det1}
\end{eqnarray}
The normalization is chosen such that 
$ \lim_{B\rightarrow 0} F(B,R|\alpha)=0.  $
The free energy is then $F=v^{-1}\zeta'(0)$.

In the zeta-regularised determinant an arbitrary
scale $\mu$ appears and it is convenient to work with scaled variables
\begin{eqnarray*}
\beta=\sqrt{2\hat B}/\mu, \ \rho=\mu R, \ \xi=\lambda(B)/\mu, \ 
\xi_0=\lambda(0)/\mu, \ 
\end{eqnarray*}
and where the dimensionless quantity $z=BR^2/2=\beta^2\rho^2/4$
appears prominently. Moreover it is convenient
to analytically continue $\alpha \rightarrow -i \vartheta -\pi/2$
to guarantee a real spectrum of eigenvalues of the Dirac
operator for all $\vartheta$. We shall regard the background domain field
as being self-dual in the following. Then, 
Eqs.(\ref{L-+},\ref{L--}) can be recast into a form
determining the rescaled eigenvalues $\xi$, namely
\begin{eqnarray}
M(k+2-\xi^2/\beta^2,k+2,z)
+e^\vartheta\frac{\beta^2\rho}{2\xi}
\left[M(k+2-\xi^2/\beta^2,k+2,z)
-\frac{k+2-\xi^2/\beta^2}{k+2}
M(k+3-\xi^2/\beta^2,k+3,z)\right]=0 \ \ \
\label{mppm}
\end{eqnarray}
for $\xi^{-+}$ and
\begin{eqnarray}
\label{pmpm}
M(-\xi^2/\beta^2,k+1,z)+
e^{\vartheta}\frac{\xi\rho}{2(k+1)}M(1-\xi^2/\beta^2,k+2,z)=0
\end{eqnarray}
for $\xi^{--}$.
The zeta function $\zeta(s)$ breaks up into
two parts \cite{DGS98},
respectively symmetric (S) and antisymmetric (AS) with respect to 
$\xi\to-\xi$   
\begin{eqnarray*}
\zeta(s)=\zeta_{\rm S}(s)+\zeta_{\rm AS}(s),
\end{eqnarray*}
with
\begin{eqnarray}
\zeta_{\rm S}(s)& =& \frac{1}{2}\left(1+e^{\mp i\pi s}\right)
\zeta_{\!\not\!D^2}(s/2), \label{zetaS} \\
\zeta_{\rm AS}(s)&=&\frac{1}{2}\left(1-e^{\mp i\pi s}\right)\eta(s).
\label{zetaAS}
\end{eqnarray}
The assumption behind these representations is that the spectrum
of eigenvalues can be well ordered according to the magnitude
of the eigenvalues, $|\xi|$, which is
certainly the case with the solutions to Eqs.(\ref{mppm},\ref{pmpm}). 
However there is an ambiguity
which can be fixed by specifying whether the smallest-in-magnitude
eigenvalue is either positive or negative. This determines
respectively the sign choice in Eqs.(\ref{zetaS},\ref{zetaAS}). 

Thus the key quantities devolve into the zeta function for the
squared Dirac operator and the asymmetry function respectively:
\begin{eqnarray}
\zeta_{\!\not\!D^2}(s)&=&\sum_{k,n,\kappa}(k+1)
\left(\frac{1}{[\xi^\kappa_{kn}(B)]^{2s}}
 -
\frac{1}{[\xi^\kappa_{kn}(0)]^{2s}}
\right)
\label{zeta1}
\\
\eta(s)&=&\sum_{k,n,\kappa}(k+1)\left(
\frac{{\rm sgn}(\xi^\kappa_{kn}(B))}{|\xi^\kappa_{kn}(B)|^{s}}
-\frac{{\rm sgn}(\xi^\kappa_{kn}(0))}{|\xi^\kappa_{kn}(0)|^{s}}
\right).
\label{eta1}
\end{eqnarray}
It should be stressed that
in the presence of bag-like boundary conditions
$\zeta_{\rm S}$ and $\zeta_{\rm AS}$ do not have the meaning of 
parity conserving and parity violating terms
since a parity transformation in terms of eigenvalues is given by 
$\xi(\vartheta) \to -\xi(-\vartheta)$
and both spectral functions contain parity conserving and violating terms.
Thus the determinant for a given parameter $\vartheta$ is defined by
\begin{eqnarray}
\zeta'(0)=\left(\frac{1}{2}\zeta'_{\!\not\!D^2}(0)
\pm i\frac{\pi}{2}\zeta_{\!\not\!D^2}(0)\mp i\frac{\pi}{2}\eta(0)\right),
\label{zetaprime}
\end{eqnarray}
with the normalization chosen in Eq.(\ref{det1}) such that $\zeta'(0)$ vanishes
as $B\rightarrow 0$.

Spectral sums over quantum labels $N$
can be computed using a representation of the
sum as a contour integral of the logarithmic derivative
of the function whose zeroes determine the spectrum, 
\begin{equation}
\sum_{N} {1 \over {\xi_N^s}} = \frac{1}{2\pi i}
\oint_{\Gamma} \frac{d\xi}{\xi^s} \frac{d}{d\xi}\ln f(\xi),
\label{intrep}
\end{equation}
see~\cite{intformula},
where the zeroes of $f(\xi)=0$ are $\xi=\xi_N$ and the contour is
chosen such that all zeroes are enclosed. With real parameter
$\vartheta$, the poles lie on the real
axis, and there is no pole at the origin for any $\vartheta$. 
By deforming the contour and accounting for the vanishing of
contributions to the integral at infinity, the expressions arising from 
Eq.~({\ref{intrep}) can be transformed into real integrals. 
The following representations
for the two spectral functions are eventually obtained
\begin{eqnarray}
\zeta_{\not D^2}(s)&=&\rho^{2s}\frac{\sin(\pi s)}{\pi}
\sum_{k=1}^\infty k^{1-2s} 
\int_0^\infty \frac{dt}{t^{2s}}\frac{d}{dt}\Psi(k,t|z,\vartheta),
\nonumber\\
\eta(s)&=& \rho^{s}\frac{\cos(\pi s/2)}{i\pi}
\sum_{k=1}^\infty k^{1-s} 
\int_0^\infty \frac{dt}{t^{s}}\frac{d}{dt}\Phi(k,t|z,\vartheta),
\label{etas}
\end{eqnarray}
with $\Psi$ and $\Phi$ being the sum of contributions from the
two polarisations, taking the form
\begin{eqnarray*}
\Psi(k,t|z,\vartheta) &= & \sum_{\kappa=\pm}\ln \left(
{ {A_{\kappa}^2(k,t|z)+e^{2\vartheta} B_{\kappa}^2(k,t|z)}} \over
{A^2(k,t) + e^{2\vartheta} B^2(k,t)} \right)
\nonumber \\
\Phi(k,t|z,\vartheta) &=& \sum_{\kappa=\pm}\ln \left(
{ {A_{\kappa}(k,t|z)+i e^{\vartheta} B_{\kappa}(k,t|z)} \over
{A_{\kappa}(k,t|z) -i e^{\vartheta} B_{\kappa}(k,t|z)}} \right)
\nonumber 
\end{eqnarray*}
and where 
\begin{eqnarray}
\label{AB}
A_{-}(k,t|z)&=&M\left(\frac{k^2t^2\rho^2}{4z},k+1,z\right), 
 \nonumber\\
B_{-}(k,t|z)&=&\frac{kt\rho}{2(k+1)}
M\left(1+\frac{k^2t^2\rho^2}{4z},k+2,z\right),
\nonumber\\
A_{+}(k,t|z)&=&M\left(-\frac{k^2t^2\rho^2}{4z},k+1,-z\right), 
\nonumber\\
B_{+}(k,t|z)&=&\frac{2z}{kt\rho}
\left[M\left(-\frac{k^2t^2\rho^2}{4z},k+1,-z\right)
-\frac{k+1+\frac{k^2t^2\rho^2}{4z}}{k+1}
M\left(-\frac{k^2t^2\rho^2}{4z},k+2,-z\right)\right],
\nonumber\\
A(k,t)&=&\frac{2^{k}k!}{(kt\rho)^{k}}I_k(kt\rho), \nonumber \\
B(k,t)&=&\frac{2^{k}k!}{(kt\rho)^{k}}I_{k+1}(kt\rho).
\end{eqnarray} 
The Bessel functions $I_k$ emerge from the limit $B\rightarrow 0$
with the normalization of the determinant as specified above.
The next step is to expand the confluent hypergeometric functions
in $1/k$, similar to the Debye expansion of Bessel functions, for example
\begin{eqnarray}
M\left(\frac{k^2t^2\rho^2}{4z},k+1,z\right) &=& 
C(t\rho,k)\sum_{n=0}^\infty {{M_n(t\rho,z)}\over k^n}.
\label{Mexpand}
\end{eqnarray}
The form of the prefactors and the $M_n(x,z)$
functions are given in Appendix A for the various Kummer
functions appearing in the $A_{\kappa}$ and $B_{\kappa}$.

Before proceeding with more detail, let us give an
overview of the subsequent steps. 
The expansions Eq.~({\ref{Mexpand}) are first
inserted in $\zeta(s)$ and $\eta(s)$. Then
the integrals over $t$ can be evaluated term by term
in the series in $1/k$. The order of summation over $k$ and $n$ 
are then exhanged. This is the most subtle step, which
we discuss below. But it means that now the
sums over $k$ can be read off in terms of the 
Riemann zeta function.  
The resulting $\zeta_{\not D^2}(s)$ then has the structure
\begin{eqnarray}
\zeta_{\not D^2}(s)=s \rho^{2s} {\sin(\pi s)\over i\pi}
\sum_{n=0}^\infty \zeta_{\rm R}(2s+n-1)f(z^2|n) 
+\delta \zeta_{\not D^2}(s),
\label{struct}
\end{eqnarray} 
where the term $\delta \zeta$ denotes 
those potentially present contributions coming from 
interchange of the order of summations over $n$ and $k$.
A similar structure appears for the asymmetry spectral function as well.
Then we are interested in the decomposition of the
resulting expressions around $s=0$.
In this limit the first term in Eq.(\ref{struct}) provides only a contribution
from the $n=2$ term and can be calculated with relative ease 
since only the lowest coefficients $M_1$ and $M_2$
in Eq.~(\ref{Mexpand}) contribute.  However, as occurs
in even simple problems \cite{Elizbook}, 
the second term in Eq.(\ref{struct}) is much more difficult to
compute. To achieve this one needs to know coefficients $f(z^2|n)$ as a 
function of  continuous variable $n$. 
Below we will elucidate on the contribution
from the first term alone, bearing in mind 
the necessity of a complete analysis. 

Now in more detail, using the expansions of the type Eq.(\ref{Mexpand}) 
as given in Appendix A, we arrive at the following expansions of $\Psi$
and $\Phi$ in $1/k$:
\begin{eqnarray*}
\Psi(k,t|z,\vartheta)&=&z^2\left[\frac{\Psi_1(y)}{k}
+\frac{\Psi_2(y|\vartheta)}{k^2}\right] + O(1/k^3),
\\
\Phi(k,t|z,\vartheta)&=&z^2\frac{\Phi_2(y|\vartheta)}{k^2} + O(1/k^3),
\end{eqnarray*}
with the coefficient functions
\begin{eqnarray*}
\Psi_1(y)&=&-\frac{2}{3}\frac{y(y-1)(3y+1)}{(y+1)^2},
\\
\Psi_2(y|\vartheta)&=&
\frac{y^2(y-1)}{\left[y+1-(y^2-1)^2e^{2\vartheta}\right]^2}      
\\
&&\times\left[(y^2-1)(2y^3+6y^2+7y+1)-2(y+1)(2y^4+5y^3+5y^2-y+1)e^{2\vartheta}
\right.\nonumber \\
&&\left. +(y-1)(2y^4+6y^3+5y^2-4y-1)e^{4\vartheta}\right],
\nonumber \\
\Phi_2(y|\vartheta)&=&-2ie^{\vartheta}y^2\sqrt{1-y^2}
\frac{(1-y)^2e^{2\vartheta}-(1+y)^2}{\left[1+y+(1-y)e^{2\vartheta}\right]^2},
\end{eqnarray*}
with $y=\frac{1}{\sqrt{1+t^2}}$.
We thus get for $s<<1$ 
 \begin{eqnarray*}
\zeta_{\not D^2}(s)&=&z^2(1+2s\ln\rho)s\left[-\frac{1}{2}{\cal I}_1 
+\left(\frac{1}{2s}+\gamma\right){\cal I}_2) \right]
\\
{\cal I}_1&=&\int_0^\infty\frac{dt}{t^{2s}}\frac{d}{dt}\Psi_1(y)
=-s+O(s^2)
\\
{\cal I}_2&=&\int_0^\infty\frac{dt}{t^{2s}}\frac{d}{dt}\Psi_2(y|\vartheta)
=2s\left[
-\frac{1}{4}-\ln2+\ln\left(1+e^{2\vartheta}\right)
\right]+ O(s^2)
\end{eqnarray*} 
and
\begin{eqnarray*}
\eta(s)&=&-\frac{z^2}{i\pi}(1+s\ln\rho)\frac{1}{s}{\cal J}_2
\\
{\cal J}_2&=&2ie^{\vartheta}\int_0^\infty\frac{dt}{t^{s}}
\frac{d}{dt}\Phi_2(y|\vartheta)
=-s\frac{i}{2}
\left[\pi+2{\rm Arctan}({\rm sinh}(\vartheta)\right]+O(s^2)
\end{eqnarray*} 
upon performing the $t-$ (or $y-$) integrations and
where we have used that
\begin{eqnarray*}
\zeta_{\rm R}(0)=-\frac{1}{2}, \ \ \zeta_{\rm R}(1+2s)=\frac{1}{2s}
+\gamma+O(s).
\end{eqnarray*} 

The final results for the first term in Eq.~(\ref{struct})
and its analogue in the asymmetry function $\eta(s)$
are then summarised in the following equations:
\begin{eqnarray*}
\zeta_{\not D^2}(0)&=&0, \nonumber \\
\zeta'_{\not D^2}(0)&=&z^2\left[
-\frac{1}{4}-\ln2+\ln\left(1+e^{2\vartheta}\right)
\right]
\nonumber\\
\eta(0)&=&\frac{z^2}{2}+\frac{z^2}{\pi}{\rm Arctan}({\rm sinh}(\vartheta))
\end{eqnarray*} 
and hence
\begin{eqnarray}
\label{finalzeta}
\zeta'(0)&=& \frac{z^2}{2}\left[-\frac{1}{4}-\ln2\pm 
i\frac{\pi}{2}\right.
 \\
&&\left.+\ln\left(1+e^{2\vartheta}\right)
\mp i {\rm Arctan}({\sinh}(\vartheta))\right]+\delta \zeta'(0),
\nonumber
\end{eqnarray} 

Now we can straightforwardly continue 
$\vartheta\rightarrow i \alpha + i \pi/2$. 
The final result for the free energy 
$F$ is complex with the imaginary part of the
form 
\begin{equation}
\label{anomaly}
\Im{F}= \pm 2 q \rm{Arctan}(\tan(\alpha))
\end{equation}
where $q$ is the absolute value of topological charge in a domain.
This charge is not integer here in general but the anomalous term is 
$\pi n$ periodic in $\alpha$.
This is the Abelian anomaly as observed
within the context of bag-like boundary conditions by
\cite{WD94}. 
Its appearance here is in the spirit of
the derivation by Fujikawa \cite{Fuj80}, where the phase appears as
an extra contribution under a chiral transformation on the
fermionic measure of integration. 

However the analytic continuation of Eq.(\ref{finalzeta}) also 
exposes an $\alpha$ dependent real part,
\begin{eqnarray*}
F(\alpha)
=\frac{z^2}{2}\left[-\frac{1}{4}+\ln(1\pm\cos(\alpha))\right]
\end{eqnarray*}
where the signs are according to the prescription in
Eqs.(\ref{zetaS},\ref{zetaAS}). There is no reason to expect
that anything other than the anomaly should appear in
the logarithm of the determinant. Thus the likelihood is
that this additional real part contribution should be
cancelled by the contributions $\delta \zeta$ in Eq.(\ref{struct}),
which remains to be verified.

Below we assume 
that the anomaly Eq.~(\ref{anomaly}) provides the entire result 
for the $\alpha$-dependent part
of the free energy of massless fermions in a domain, which leads to 
intriguing consequences.

\subsection{Massive fermions}

Inclusion of an infinitesimally small fermion mass 
leads to a modification of the free energy by a term which is linear
in mass to leading order, namely \cite{DGS98} 
\begin{eqnarray}
F=F_{m=0} +
i\frac{m}{\mu v}\eta(1).
\label{massivefreeenergy}
\end{eqnarray}
Using again the representation Eq.(\ref{etas})
but now in the vicinity of $s=1$
one can find the following representation
for the  summed  contributions
of both polarisations in the self-dual domain
to the
asymmetry function 
\begin{eqnarray}
\label{eta-1-1}
\eta(s)=i \mu R e^{i\alpha} \frac{\cos(\pi s/2)}{\pi(1-s)}
\sum_{k=1}^\infty k^{1-s}\frac{k}{k+1}
\left[
1 + M(1,k+2,z) -\frac{z}{k+2}M(1,k+3,-z)
\right] .
\end{eqnarray}
We next evaluate the asymptotic behaviour in $k$.
A singular term
as $s\rightarrow 1$ can be extracted, which turns out to
be independent of the field (that is, $B$) 
and is canceled by the normalization. 
The final expression for the free energy density
for a self-dual domain, including the contribution linear in mass, is
\begin{eqnarray*}
F^{(\rm sd)}=2i q {\rm Arctan}(\tan(\alpha)) 
- e^{i\alpha} m  \prec\bar\psi \psi\succ,
\end{eqnarray*}
where we have used the suggestive notation
\begin{eqnarray}
\label{cond0}
\prec \bar\psi \psi\succ
=\frac{1}{\pi^2 R^3}
\sum_{k=1,z=z_1,z_1,z_2}^\infty \frac{k}{k+1}
\left[
M(1,k+2,z)-\frac{z}{k+2}M(1,k+3,z)-1
\right]
\end{eqnarray} 
coming from $\eta(1)$ with the sum over $z$ 
correponding to a colour trace.

Let us summarise the results of this section as
follows. The  $\alpha$ dependent part of the free energy of a self-dual domain 
for massive quarks is complex with the following
real and imaginary parts:
\begin{eqnarray}
F&=&\Re{F}+i\Im{F} \nonumber \\
\Re{F} &=& - m \cos \alpha \prec \bar\psi \psi\succ \label{realF} \\
\Im{F} &=& 2 \frac{q}{v}\rm{Arctan}(\tan(\alpha)) 
- m \sin \alpha \prec \bar\psi \psi\succ.
 \label{imagF}
\end{eqnarray}
The free energy of an anti-self-dual domain
is obtained via complex conjugation. 

\section{Ensemble free energy and  chiral symmetries.}

\subsection{One flavour case: quark condensate and $U_{\rm A}(1)$}

Under the assumption that only the anomalous term depends on the chiral angle, 
the part of the free energy 
density ${\cal F}$  relevant for the present consideration of
an ensemble of $N\to\infty$ domains
with both self-dual and anti-self-dual configurations takes the form
\begin{eqnarray*}
e^{-vN{\cal F}}&=&{\cal N}\prod_j^N\int_0^{2\pi}d\alpha_j \frac{1}{2}
\left[e^{iv\Im F(\alpha_j)}+e^{-iv\Im F(\alpha_j)} \right]
\nonumber \\
&=&{\cal N}\prod_j^N\int_0^{2\pi} d\alpha_j
e^{ \ln(\cos(v\Im F(\alpha_j)))}
\nonumber
\\
&=&{\cal N}\exp\left(N\max_\alpha
\ln(\cos(v\Im F(\alpha)))\right) .
\nonumber
\end{eqnarray*} 
The maxima (minima of the free energy density)
are achieved at $\alpha_1=\dots=\alpha_N=\pi n$.

In the absence of a quark mass, only the anomaly contribution
in the imaginary part, $\Im{F}$, of the free energy of a single domain
appears under the logarithm of the cosine and defines the minima of the free
energy density, $\ln(\cos(v\Im F(\alpha)))=0$.  
Thus for massless quarks there is no
continuous $U_A(1)$ symmetry in the ground states, rather
a discrete $Z_2$ chiral symmetry.
The anomaly plays a peculiar role here: selecting out those
chiral angles which minimise the free energy
so that the full $U_A(1)$ group is no longer reflected in
the vacuum degeneracy.
It should be stressed here that this residual discrete degeneracy is sufficient
to ensure a zero value for the quark condensate in the absence of a mass term
or some other external chirality violating sources.

Now switching on the quark mass, we see this discrete
symmetry spontaneously broken, and one of the two vacua selected in the
infinite volume limit according to the sign of the mass.
In this case (for these conventions of boundary
condition and mass term), the
minimum at $\alpha=0$ is selected. 
The quark condensate can be now extracted from the free energy via
\begin{eqnarray*}
\langle \bar\psi(x)\psi(x)\rangle=-\lim_{m\to0}\lim_{N\to\infty} 
(vN)^{-1} \frac{d}{dm}e^{-vN{\cal F}(m)}.
\end{eqnarray*} 
Taking the thermodynamic limit $N\to\infty$ and then $m\to0$ gives  
a nonzero condensate 
\begin{eqnarray}
\label{cond-f}
\langle \bar \psi(x) \psi(x)\rangle =-\prec \bar \psi \psi\succ.
\end{eqnarray} 
According to Eq.~(\ref{cond0}) the condensate is equal to 
\begin{eqnarray}
\label{vcond-f}
\langle \bar \psi(x) \psi(x)\rangle  = -(237.8 \ {\rm MeV})^3
\end{eqnarray} 
for the values of field strength $B$ and domain radius $R$ 
fixed earlier by consideration of the pure gluonic 
characteristics of the vacuum -- string tension, topological succeptibility 
and gluon condensate.
A nonzero condensate is generated without 
a continuous degeneracy of the ground states of the
system.

\begin{figure}[htb]
\vspace{17mm}
\includegraphics{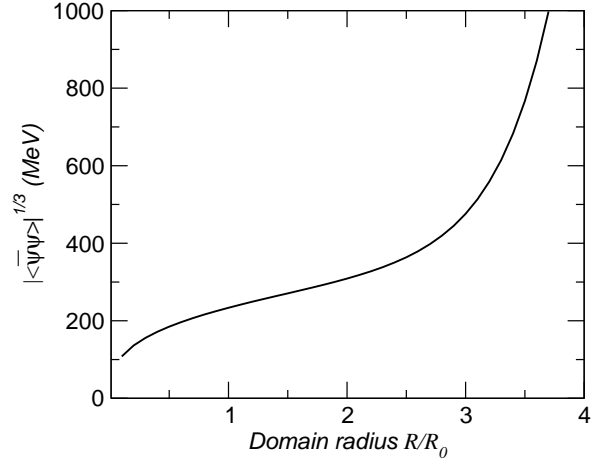}
\caption{Absolute value of quark condensate as a function
of domain radius ($R_0=(760 {\rm MeV})^{-1}$).}
\label{fig:cond}
\end{figure}

Dependence of the condensate on the domain radius $R$ is illustrated in 
Fig.\ref{fig:cond}.
As expected the condensate diverges with $R\to\infty$, since in 
this limit the number of low-lying strongly chiral modes is growing. 
We will return to the discussion of this feature again in the section on the
meson spectrum.

\subsection{Multiflavour case: spontaneous breakdown of 
$SU_{\rm L}(N_f)\times SU_{\rm R}(N_f)$}

The question remains whether any continuous directions in the space of vacua
are to be expected when the full flavour chiral symmetry is brought
into play. For this we must generalise the analysis. 
We consider $N_f$ massless quark flavours.
Firstly, we observe that the fermion boundary
condition in Eq.(\ref{quarkbc})
explicitly breaks all chiral symmetries, flavour singlet
and non-singlet (see also \cite{WD94}). 
Thus the procedure we have used here of integrating
over all $\alpha$ does not suffice to restore the full chiral
symmetry of the massless QCD action. Rather, the boundary
condition must be generalised to include flavour non-singlet
angles,
\begin{equation*}
\alpha \rightarrow \alpha +\beta^a T^a,
\end{equation*}
with $T^a$ the $N_f^2-1$ generators of $SU(N_f)$. 
Then integration over $N_f^2$ angles 
$\alpha,\beta^a \in [0,2\pi]$ must be performed
for a fully chiral symmetric ensemble. The spectrum of the Dirac
problem now proceeds quite analogously, except that the
boundary condition mixes flavour components, thus an additional
projection into flavour sectors
is required in order to extract the eigenvalue equation analogous to
Eq.(\ref{bc-main}). 

For $N_f=2$ the boundary condition can be chosen as
\begin{equation}
i\!\not\!\eta(x) 
e^{i(\alpha +{\vec \beta}\cdot{\vec {\sigma}}/2) \gamma_5}
\psi(x)=\psi(x)
\label{genquarkbc}
\end{equation}
where flavours now mix on the boundary.
We must now solve the Dirac spectrum in the
presence of this mixing. The
spinors $\chi$ and $\varphi$ used in our solutions
become now isospin doublets. We need to project
the boundary condition onto separate equations for
eigenvalues.

Things proceed much as before in the spin sector
with our previous decompositions and projectors. The 
boundary condition will devolve to the structure
\begin{equation*}
\chi[\lambda] = e^{i(\alpha +{\vec \beta}\cdot{\vec {\sigma}}/2)} 
\varphi[\lambda]
\end{equation*}
where the $\gamma_5$ in the exponent is eliminated via the projection
into chirality eigenspinors. The equation still mixes
the flavour components and cannot be solved for an eigenvalue $\lambda$.

We introduce projectors
\begin{equation*}
P_{\pm}(\beta) \equiv { {1 \pm {\hat \beta}\cdot {\vec \sigma}}\over 2}
\end{equation*}
which have the property
\begin{eqnarray*}
P_{\pm}(\beta)  e^{-i {\vec \beta}\cdot{\vec {\sigma}}/2} 
=e^{\pm i |{\vec \beta}|/2} P_{\pm}(\beta).
\end{eqnarray*}
Note here the appearance of the magnitude of
the triplet ${\vec \beta}$ in the exponent.
Projecting the spinors thereby we obtain separate equations
for eigenvalues $\lambda[\beta,\pm]$ where we suppress
all other quantum numbers previously dealt with,
\begin{equation*}
\chi[\lambda[\beta,\pm]] = e^{i(\alpha \pm  |{\vec \beta}|/2) } 
\varphi[\lambda[\beta,\pm]]
\end{equation*}

It is convenient to define two new angles,
\begin{equation*}
{\alpha}_{\pm} = \alpha \pm  |{\vec \beta}|/2
\end{equation*}
and in terms of these angles we can calculate the
quark determinant separately for each flavour projection as before.
The free energy for each flavour is now a function of four constrained angles
\begin{equation*}
F_{\pm}=F_{\pm}(\alpha,\beta_+,\beta_0,\beta_-)
\end{equation*}
where we use this suggestive notation instead of $\beta_1,\beta_2,\beta_3$,
and where ${\pm}$ denotes the isospin projection
for the two quark flavours. 
For a given self-dual domain the total free energy will be
a sum
\begin{equation*}
F_T(\alpha,{\vec \beta}) = F({\alpha}_+) + F({\alpha}_-) 
\end{equation*}
where $F$ is the result from the one-flavour case.
Under the assumption that the anomaly provides the entire
chiral angular dependence of the determinant, 
the nonsinglet angles $\beta^a$ drop out due to the form
$2q \arctan(\tan(\alpha))=2q(\alpha+2n \pi)$. 
This expresses the known result that the anomaly
only depends on the flavour singlet directions or
is Abelian. Thus for an ensemble of domains
the free energy is identical to that for one massless
flavour, namely it depends only on the Abelian angle
$\alpha$. 
Thus for $N_f=2$, 
the $U_A(1)$ direction remains fixed by energy minimisation
while the $SU(2)_L\times SU(2)_R$ directions represent degeneracies
in the space of ground states in 
the thermodynamic limit.

For $N_f=3$ the boundary condition will now involve the flavour mixing matrix
\begin{equation*}
e^{i \beta^a \lambda^a/2}
\end{equation*}
in terms of the Gell-Mann matrices.
It is now harder to explicitly diagonalise and project
into flavour sectors. Nonetheless the form of the result
is clear. The Cartan subalgebra
consists of the diagonal generators $\lambda^3$ and $\lambda^8$. Thus the
result of a diagonalisation of the argument of the exponential
will have the form,
\begin{equation*}
\beta^a \lambda^a/2 \rightarrow {b^A} \lambda^A/2, \ A=3,8
\end{equation*}
where $b^i={b^i}(\beta^a)$, functions of the original nonsinglet angles
analogous to $|{\vec \beta}|$ for $N_f=2$.
Thus diagonalisation will amount to,
\begin{eqnarray*}
e^{i \beta^a \lambda^a/2}
\rightarrow  {\rm{diag}}\left( 
e^{b^3+\frac{1}{\sqrt{3}}b^8}, 
e^{-b^3+\frac{1}{\sqrt{3}}b^8}, 
e^{-\frac{2}{\sqrt{3}}b^8}
\right)
\end{eqnarray*}
So projection of the combined $U(1)\times SU(3)$ chiral
boundary condition will lead to three
sets of equations with the eigenvalues depending on
the combinations of angles
\begin{eqnarray*}
B_1 & = & \alpha + b^3+\frac{1}{\sqrt{3}}b^8, \nonumber \\
B_2 & = & \alpha - b^3+\frac{1}{\sqrt{3}}b^8, \nonumber \\
B_3 & = & \alpha - \frac{2}{\sqrt{3}}b^8
\end{eqnarray*}
where evidently
\begin{equation*}
\sum_i B_i = 3 \alpha
\end{equation*}
reflecting ultimately the tracelessness of the generators.
The free energy for a given domain will be the sum of three terms
\begin{equation*}
F_T = \sum_{i=1}^3 F(B_i)
\end{equation*}
with $F$ being the same expression evaluated for the one-flavour case.
Once again, the assumption of the anomaly means cancellation of
the $\beta^a$ dependent functions $b^A$ leaving only the $\alpha$
dependence in the free energy.
Once again energy minimisation constraints the $U_A(1)$ direction,
leaving the $SU(3)_L\times SU(3)_R$ directions unconstrained.
Thus for $N_f=3$ one expects eight, not nine, continuous
directions in the space of vacua.

The argument for arbitrary $N_f$ is now evident, the key ingredients
being the tracelessness of $SU(N_f)$ generators and the specific form
of the anomaly for the quark determinant.

\section{Estimation of meson masses.}

The consequences of the above realisation of chiral symmetries
should be seen in the meson spectrum providing the splitting between 
pseudoscalar and vector mesons and between the 
pseudoscalar octet and $\eta'$.

The source for the difference between the masses of the
octet states and $\eta'$  
can be recognised in the drastic difference 
between correlators of the
flavour octet $J^{a}_{\rm P}(x)$ and 
singlet $J_{\rm P}(x)$ pseudoscalar
quark currents as they appear in the domain model,
\begin{eqnarray}
\langle  J^{a}_{\rm P}(x) J^{b}_{\rm P}(y)\rangle
&=&\overline{\langle \langle J^{a}_{\rm P}(x) J^{b}_{\rm P}(y)\rangle\rangle}
\nonumber\\
\langle  J_{\rm P}(x) J_{\rm P}(y)\rangle
&=&\overline{\langle \langle J_{\rm P}(x) J_{\rm P}(y)\rangle\rangle}
-
\overline{\langle \langle J_{\rm P}(x)\rangle\rangle \langle\langle 
J_{\rm P}(y)\rangle\rangle}.
\label{sv-corr}
\end{eqnarray}
Here double brackets denote integration
over quantum fluctuation fields and the overline means
integration over all configurations in the domain ensemble.
The second term in the RHS 
of the flavour singlet correlator contains two quark loops and 
is  subleading in $1/N_c$ compared with the first one-loop contribution.
The second term is entirely determined by the correlation function of the 
background gluon field $B$ in the ensemble (\ref{corfunctn}) and is 
proportional to the quark condensate squared. 
The analogous two-loop term in the  flavour nonsinglet correlator
is equal to zero due to the trace over flavour indices.
It should be added that the pseudoscalar condensate,
$\langle \bar{\psi} \gamma_5 \psi \rangle$, naturally vanishes 
since parity is not broken in the ensemble of domains.
Thus massless modes can be expected in the  nonsinglet
channel, but not in the flavour singlet due to the 
additional term in the correlator. This general structure of correlators is 
exactly the same as in the 
instanton liquid model~\cite{Shuryak} and manifests
the mechanism for eta-prime mass generation
proposed by Witten in \cite{largeNc} and appreciated in chiral
perturbation theory by \cite{chpert}.  
The correspondence between domain model correlators to 
original QCD diagrams is illustrated
in Fig.~\ref{fig:diagr}.  The grey background denotes averaging over 
the domain ensemble and is intended to represent the nonperturbative 
intemediate range part of 
the gluon exchange in the original
QCD diagrams (a') and (b').  Diagrams (a) and (b) correspond to the 
first and second terms in the RHS of Eq.~(\ref{sv-corr}). 
Averaging in diagram (b) relates to
both quark loops, which is the domain model
representation
of the exchange between quark loops in (b') by infinitely many gluons in the 
original QCD representation. It should be noted that diagrams (a) and (b) 
in Fig.~\ref{fig:diagr} 
are the lowest order contributions in the fluctuation gluon fields $Q$ 
which are treated as perturbations of the background. Higher orders include
exchange by gluonic fluctuations in the presence of the domain mean field.

\begin{figure}[htb]
\vspace{17mm}
\includegraphics{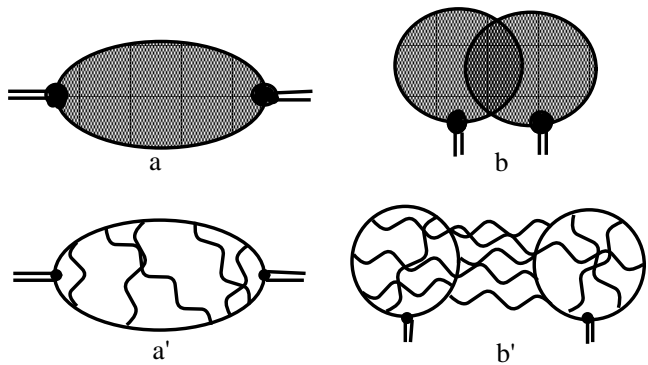}
\caption{Diagrammatic representation of two types of contributions to the 
flavour singlet correlator: (a) and (b) as in the domain model;
(a') and (b') as in full QCD. The
grey background in (a,b) represents the nonperturbative part of the 
gluon exchange indicated in diagrams (a',b').}
\label{fig:diagr}
\end{figure}
Direct calculation of meson masses within the domain model is not available 
yet. However we can estimate the effect of contributions subleading in $1/N_c$
in the flavour singlet pseudoscalar correlator within a calculational scheme 
which is quite close to the domain model in this respect. 
This is the main purpose of this section.
Simultaneously we will schematically expose the form of the effective
action for collective colourless meson-like modes as is expected to emerge 
in the domain model and estimate the value of the typical dimensionless 
parameter $BR^2$  but now from the meson spectrum. 
The model which will be used for this purpose is based on the bosonisation
of a one gluon exchange interaction between quark currents and in which both 
quark and gluon propagators are exact solutions in the presence of a 
background (anti-)self-dual homogeneous gluon field 
as in the papers ~\cite{PRD95,PRD96,pf}, but here with additonal
correlations of the type (b) in Fig.~\ref{fig:diagr}. 
In~\cite{PRD95,PRD96,pf}, as well as the formulation of the model 
based on a homogeneous gluon field,
applications to the calculation of the spectrum of light mesons, 
their orbital excitations,  heavy-light mesons, heavy quarkonia, 
decay constants and form-factors are given. 

In the presence of a homogeneous (anti-)self-dual background field the 
quark condensate emerges due to chiral zero modes,
but, since there is  a continuum of such  modes in the homogeneous field,  
``an overkill'' occurs~\cite{Smilga,EfNed}:
$$m\langle\bar \psi\psi\rangle  \propto - B^2  , \ \ m\to 0,$$
and the limit $m\to0$ cannot be defined properly. However,
due to the same zero modes the momentum representation correlators of 
scalar, pseudoscalar, vector and axial vector currents have 
the following limits for $m\to0$
\begin{eqnarray}
&&\tilde\Pi_{P/S}\to\mp \frac{B^2}{m^2}F_1(p^2/B), 
\nonumber\\
&&\tilde\Pi_{V/A}\to\mp BF_2(p^2/B).
\label{PSVA}
\end{eqnarray}
Such a qualitatively different behaviour of correlators becomes manifest 
already at $m^2/B\sim O(1)$ and leads to a strong splitting of the masses
of the corresponding mesons ensuring light pions and 
heavy $\rho$-mesons in particular~\cite{PRD96} (see also
\cite{Shuryak}).  The fitted values of quark masses in the model 
correspond to the constituent masses.

The definition of the chiral limit in that model 
thus cannot be given in terms of quark masses.
However, as we discuss below
it is nonetheless possible to define a
regime when the calculated pion mass vanishes.
The curious side of this approach
to the description of chiral properties of the light mesons  
is that the  zero-mode mechanism of condensate formation alone is exploited
without any use made of a Nambu-Jona-Lasinio or, more generally,
Dyson-Schwinger equation type mechanism~\cite{DeGrand}. 
The inclusion in this model of the additional correlations (b,b') 
of Fig.~\ref{fig:diagr} for the flavour singlet 
channel enables this approach to also work phenomenologically for 
$\eta-\eta'$ masses with the chiral limit so-defined. 

In the domain model studied in the first
part of this paper 
the quark condensate appears also due to 
specific chiral properties of  (now nonzero) Dirac modes 
in gluon (domain-like) background fields 
and not due to a four-fermion 
interaction. The chiral properties of these low-lying modes 
resemble the properties  of
zero modes in the homogeneous-field model.
The condensate diverges as $R\to\infty$ as shown in 
Fig.~\ref{fig:cond} much like the condensate in the homogeneous
field model diverges as $1/m$. 
We expect that this dependence
on $R$ in the domain model should have a similar effect on meson 
correlators and masses as the $1/m$ singularity does in the 
homogeneous field case.
The characteristic scale of higher modes is defined by the field strength 
$B$ rather than the domain radius for $BR^2\gg1$.  
Thus one would expect that the heuristic consideration below 
should be more consistent with the domain model picture if $BR^2\gg1$.

The difference is that in the domain model flavour chiral symmetry is broken
spontaneously, which is not the case in the homogeneous field model. 
So we should expect a massless pion for massless quarks: 
the current mass and the chiral limit are well-defined,
as has been discussed in previous sections.
However a detailed description of the emergence of Goldstone modes 
is a question of detailed study of correlators and the bound state 
problem in the domain model which is still beyond our efforts.

To conclude these preliminary comments,
the direct purpose of the following calculations is to demonstrate 
the $(\pi,K)-\eta-\eta'$ 
splitting due to correlations Fig.\ref{fig:diagr}b. An auxiliary purpose,
justified by the similarity of chiral properties of zero modes in the
homogeneous model and nonzero eigenmodes in the domain model, 
is a demonstration 
that the pseudoscalar singlet-octet mass splitting  occurs simultaneously
with a correct description of pseudoscalar-vector meson splitting.

\begin{figure}[htb]
\vspace{17mm}
\includegraphics{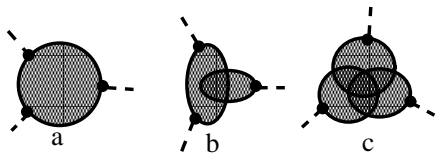}
\caption{Diagrammatic representation of the three-point vertex function 
$\Gamma^{(3)}_{{\cal Q}_1{\cal Q}_2{\cal Q}_3}$. 
Diagram (a) corresponds to the first term in
the RHS of Eq.~(\ref{gam3}), (b) and (c) to the second and third terms 
respectively. Solid lines are quark propagators $S$ in the background field, 
vertices correspond to nonlocal quark-meson vertices 
$V_{{\cal Q}_i}$ and the dashed lines represent meson fields
$\Phi_{{\cal Q}_i}$.}
\label{fig:diagr2}
\end{figure}

\subsection{Effective action for colourless composite  fields.}

We shall estimate meson masses within 
the model described by the following 
partition function 
\begin{eqnarray}
\label{PF}
&&Z={\cal N}\lim\limits_{V\to\infty}
\int D\Phi_{\cal Q}
\exp\left\{-\frac{B}{2}\frac{h^2_{\cal Q}}{g^2 C_{\cal Q}}\int dx 
\Phi^2_{\cal Q}(x)
-\sum\limits_k\frac{1}{k}W_k[\Phi]\right\},
\\
\label{mass-eq}
&&1=
\frac{g^2C_{\cal Q}}{B}\tilde \Gamma^{(2)}_{\cal QQ}(-M^2_{\cal Q}|B),
\\
\label{hmqq}
&&h^{-2}_{\cal Q}=
\frac{d}{dp^2}\tilde\Gamma^{(2)}_{\cal QQ}(p^2)|_{p^2=-M^2_{\cal Q}}.
\end{eqnarray}
The effective action in Eq.~(\ref{PF}) is expressed in terms of 
colourless composite meson fields $\Phi_{{\cal Q}}(x)$  with 
the mass $M_{\cal Q}$ defined by Eq.~(\ref{mass-eq}), where the
condensed index ${\cal Q}$ denotes isotopic and space-time indices as well 
as all possible mesonic quantum numbers (iso-spin, spin-parity in 
the ground state, total momentum, radial quantum number),
and  $k$-point nonlocal vertices 
$\Gamma^{(k)}_{{\cal Q}_1\dots{\cal Q}_k}$
\begin{eqnarray}
\label{gam2}
W_k[\Phi]&=&
\sum\limits_{{\cal Q}_1\dots{\cal Q}_k}h_{{\cal Q}_1}\dots h_{{\cal Q}_k}
\int dx_1\dots\int dx_k
\Phi_{{\cal Q}_1}(x_1)\dots \Phi_{{\cal Q}_k}(x_k)
\Gamma^{(k)}_{{\cal Q}_1\dots{\cal Q}_k}(x_1,\dots,x_k|B),
\nonumber\\
\label{gam1}
\Gamma^{(1)}_{{\cal Q}_1}&=&\overline{G^{(1)}_{{\cal Q}_1}},
\\
\Gamma^{(2)}_{{\cal Q}_1{\cal Q}_2}&=&
\overline{G^{(2)}_{{\cal Q}_1{\cal Q}_2}(x_1,x_2)}-
\Xi_2(x_1-x_2)\overline{G^{(1)}_{{\cal Q}_1}G^{(1)}_{{\cal Q}_2}},
\\
\label{gam3}
\Gamma^{(3)}_{{\cal Q}_1{\cal Q}_2{\cal Q}_3}&=&
\overline{G^{(3)}_{{\cal Q}_1{\cal Q}_2{\cal Q}_3}(x_1,x_2,x_3)}-
\frac{3}{2}\Xi_2(x_1-x_3)
\overline{G^{(2)}_{{\cal Q}_1{\cal Q}_2}(x_1,x_2)
G^{(1)}_{{\cal Q}_3}(x_3)}
\nonumber\\
&+&
\frac{1}{2}\Xi_3(x_1,x_2,x_3)
\overline{G^{(1)}_{{\cal Q}_1}(x_1)G^{(1)}_{{\cal Q}_2}(x_2)
G^{(1)}_{{\cal Q}_3}(x_3)},
\\
\label{gam4}
\Gamma^{(4)}_{{\cal Q}_1{\cal Q}_2{\cal Q}_3{\cal Q}_4}&=&
\overline{G^{(4)}_{{\cal Q}_1{\cal Q}_2{\cal Q}_3{\cal Q}_4}
(x_1,x_2,x_3,x_4)}-
\frac{4}{3}\Xi_2(x_1-x_2)
\overline{G^{(1)}_{{\cal Q}_1}(x_1)
G^{(3)}_{{\cal Q}_2{\cal Q}_3{\cal Q}_4}(x_2,x_3,x_4)}
\nonumber\\
&-&
\frac{1}{2}\Xi_2(x_1-x_3)
\overline{G^{(2)}_{{\cal Q}_1{\cal Q}_2}(x_1,x_2)
G^{(2)}_{{\cal Q}_3{\cal Q}_4}(x_3,x_4)}
\nonumber\\
&+&
\Xi_3(x_1,x_2,x_3)
\overline{G^{(1)}_{{\cal Q}_1}(x_1)G^{(1)}_{{\cal Q}_2}(x_2)
G^{(2)}_{{\cal Q}_3{\cal Q}_4}(x_3,x_4)}
\nonumber\\
&-&\frac{1}{6}
\Xi_4(x_1,x_2,x_3,x_4)
\overline{G^{(1)}_{{\cal Q}_1}(x_1)G^{(1)}_{{\cal Q}_2}(x_2)
G^{(1)}_{{\cal Q}_3}(x_3)G^{(1)}_{{\cal Q}_4}(x_4)},
\end{eqnarray}
and analogous expressions for the higher vertices.
Defining the meson-quark coupling constants $h_{\cal Q}$  by 
Eq.~(\ref{hmqq}) provides for the correct residue of the meson propagators 
at the poles and is known as a compositeness condition~\cite{compcond}.

The vertices $\Gamma^{(k)}$ are expressed {\it via} quark 
loops $G^{(n)}_{\cal Q}$  with $n$ quark-meson vertices 
\begin{eqnarray}
\label{gc}
&&\overline{G^{(k)}_{{\cal Q}_1\dots{\cal Q}_k}(x_1,\dots,x_k)}
=\int\limits_\Sigma d \sigma_j
{\rm Tr}V_{{\cal Q}_1}(x_1|B^{(j)})S(x_1,x_2|B^{(j)})\dots
V_{{\cal Q}_k}(x_k|B^{(j)})S(x_k,x_1|B^{(j)})
\nonumber\\
&&\overline{G^{(l)}_{{\cal Q}_1\dots{\cal Q}_l}(x_1,\dots,x_l)
G^{(k)}_{{\cal Q}_{l+1}\dots{\cal Q}_k}(x_{l+1},\dots,x_k)}
=\int\limits_\Sigma d \sigma_j
\nonumber\\
&&\times
{\rm Tr}\left\{
V_{{\cal Q}_1}(x_1|B^{(j)})S(x_1,x_2|B^{(j)})\dots
V_{{\cal Q}_k}(x_l|B^{(j)})S(x_l,x_1|B^{(j)})
\right\}
\nonumber\\
&&\times
{\rm Tr}\left\{
V_{{\cal Q}_{l+1}}(x_{l+1}|B^{(j)})S(x_{l+1},x_{l+2}|B^{(j)})\dots
V_{{\cal Q}_k}(x_k|B^{(j)})S(x_k,x_{l+1}|B^{(j)})
\right\},
\end{eqnarray}
where bar denotes integration  over all
configurations of the background field with measure $d\sigma_j$. 
The quark propagator 
\begin{eqnarray}
&&S(x,y)=\exp\left(-{i\over 2}x_\mu \hat B_{\mu\nu}y_\nu\right)H(x-y),
\nonumber\\
&&\tilde H(p)=\frac{1}{2v\Lambda^2}\int\limits_0^1ds
e^{-p^2/2v\Lambda^2}\left(\frac{1-s}{1+s}\right)^{m^2/4v\Lambda^2}
\nonumber\\
&&\times\left[p_\alpha\gamma_\alpha 
\pm is\gamma_5\gamma_\alpha f_{\alpha\beta}p_\beta
+m\left(P_{\pm}+P_{\mp}\frac{1+s^2}{1-s^2}
-\frac{i}{2}\gamma_\alpha f_{\alpha\beta}\gamma_\beta
\frac{s}{1-s^2}\right)\right]. 
\nonumber
\end{eqnarray}
is the exact Dirac propagator in the presence of the (anti-)self-dual
homogeneous field
\begin{eqnarray}
\label{b-eta}
\hat B_\mu(x)=-\frac{1}{2}\hat n B_{\mu\nu}x_\nu,
\
\hat B_{\mu\nu}\hat B_{\mu\rho}=4v^2\Lambda^4\delta_{\nu\rho},
\nonumber\\
f_{\alpha\beta}=\frac{\hat n}{v\Lambda^2}B_{\mu\nu}, \ 
v={\rm diag}(1/6,1/6,1/3),
\nonumber
\end{eqnarray}
where we introduced a scale $\Lambda$ related to the field strength
$B$ as 
$$
\Lambda^2=\frac{\sqrt{3}}{2}B.
$$ 

For arbitrary orbital momentum $l$  and radial quantum number $n$ 
the vertex function (more precisely -- vertex differential operator) is 
factorised  into a radial part
\begin{eqnarray}
\label{xi}
&&F_{nl}(s)=\int\limits_0^1dtt^{l+n}e^{st},
\
s=\stackrel{\leftrightarrow}{\nabla}^2\! /\Lambda^2,
\nonumber
\\
&&\stackrel{\leftrightarrow}{\nabla}_{ff'}=
\xi_f\stackrel{\leftarrow}{\nabla}-\xi_{f'}\stackrel{\rightarrow}{\nabla},
\
\xi_f=m_f/(m_f+m_{f'}),
\nonumber\\
&&\stackrel{\leftarrow}{\nabla}_\mu=\stackrel{\leftarrow}{\partial}_\mu+iB_\mu,
\
\stackrel{\rightarrow}{\nabla}_\mu=\stackrel{\rightarrow}{\partial}_\mu-iB_\mu,
\end{eqnarray}
and angular part. In our particular calculation below we will not deal with 
the excited states. For the explicit form of the angular part of the vertex 
the reader is referred to~\cite{PRD96}, where 
technical details of the derivation of meson-quark vertices 
$V_{{\cal Q}_k}(x_k|B^{(j)})$ 
and discussion of  approximations and assumptions behind the derivation
of the effective action for composite fields can be found.   
A simplified scalar field model allowing exact implementation of the method 
has been considered, and a variational 
procedure for approximate solution of  realistic problems has been formulated 
in~\cite{EfGanb}. The relation of this method to the Bethe-Salpeter equation 
and the nonrelativistic limit are analysed in~\cite{EfGanb,EfBSE}. 

Thus the $n$-point vertex $\Gamma^{(n)}$
includes contributions of two types -- the usual one-loop contributions 
(averaged over the background field) as diagram (a) in Fig.~\ref{fig:diagr2} 
and products of two or more such one-loop terms simultaneously averaged over 
the background field and multiplied by the corresponding
correlators $\Xi$  of the background field, for example as in
diagrams (b) and (c) of Fig.~\ref{fig:diagr2} for the three-point vertex. 
In the absence of a background field
these additional terms would be just disconnected diagrams and would not 
appear in the effective action.
For the purely constant field $B$ they would correspond to infinite length 
correlations breaking the cluster decomposition property in the 
effective action ($\Xi_n\equiv 1$).
The idea of domains is implemented in the above expressions by means of 
finite length correlations of the background field which ensures 
cluster decomposition.

The $n$-point correlation functions $\Xi_n$ of the background field
are defined in Eq.~(\ref{corfunctn}).
In particular the two-point correlator which will be used below
can be written in explicit form
\begin{eqnarray}
\label{corfunct}
\Xi_2(x-y)&=&\frac{N}{V}\int\limits_{V}dz \theta(x-z)\theta(y-z)
=\frac{2}{3\pi}\phi\left(\frac{(x-y)^2}{4R^2}\right),
\nonumber\\
\phi(\rho^2)&=&
\left[\frac{3\pi}{2}-3 \arcsin(\rho)-3\rho\sqrt{1-\rho^2}
-2\rho(1-\rho^2)\sqrt{1-\rho^2}\right].
\end{eqnarray}
Geometrically, the correlation function $\Xi_2(x-y)$
is equal to the volume of overlap between spherically symmetric 
four-dimensional regions determined by the characteristic 
function $\theta$ and central 
points $x$, $y$, normalized to the volume of a single such region.

Now all the elements of the effective action Eq.~(\ref{PF}) are fixed:
nonlocal meson-quark vertices $V_{{\cal Q}_1}(x_1|B^{(j)})$ and 
quark propagators  $S(x_k,x_1|B^{(j)})$ and background field correlators 
are given in explicit analytical form.

It should be noted here that meson-quark vertices are determined by the 
gluon propagator in the presence of the background 
field.   In momentum representation both quark propagator, 
meson-quark vertices 
and correlators of the background field are entire functions 
in the complex momentum plane, which means that 
quarks and gluons are absent in the model as asymptotic states, 
which is treated as colour confinement.
In the momentum representation the vertices $\Gamma^{(n)}$
have no imaginary parts and the probability of a meson  to decay  
into a quark-antiquark pair is zero.
As is shown in~\cite{PRD95,PRD96} within the model formulated above and 
clarified in a simplified exactly solvable model in~\cite{EfGanb} 
entire propagators and vertices lead to a Regge spectrum of excited
bound states represented by composite fields $\Phi_{\cal Q}$.

The only free parameters of the model are the same as in the 
domain model discussed in the first part
of this paper: the background field strength $B$, mean domain radius $R$,  
strong coupling constant $g$ and quark masses $m_f$. 
The factor $C_{\cal Q}$ in front of the 
gauge coupling constant $g$ in Eqs.(\ref{PF}) and (\ref{mass-eq}) is 
known explicitly
$$
C_{Jnl}=C_J\frac{l+1}{2^l n!(l+n)!}, \ \ C_{S/P}=\frac{1}{9}, \ \ 
C_{S/P}=\frac{1}{18}.
$$
The only place where the gauge coupling constant $g$ enters this scheme is  
the equation for meson masses Eq.~(\ref{mass-eq}); the rest of the 
effective action contains the meson-quark coupling constants calculated by 
means of Eq.~(\ref{hmqq}). 

In the next subsection we present results for the masses of  
pseudoscalar and vector nonets 
coming straightforwardly from this formalism with a special emphasis 
on the $\eta$ and $\eta'$.

\subsection{The masses and decay constants of $\eta$ and $\eta'$.}

With this representation, the calculation of masses  
of light mesons  and weak decay constants, as
performed in~\cite{PRD96},
is modified only with respect to the $\eta$ and $\eta'$ masses. 
Equations for these masses  contain terms with
additional correlation $\Xi_2$, and this splits them from the remaining light
pseudoscalar mesons.

As follows from Eqs.~(\ref{mass-eq}), the second term of (\ref{gam2}) 
and (\ref{gc}), the simplest quark loops  relevant to the 
calculation of masses are the constants     
\begin{eqnarray}
\label{j08}
&&G^{(1)}_{aP}=-{\rm Tr}\lambda^a i\gamma_5 
F_{00}(x|B^{(j)})S(x,x|B^{(j)}),
\end{eqnarray}
corresponding to the quark loops in Fig.\ref{fig:diagr}b. 
Here $a$  is the flavour index ($a=0,\dots, 8$). 
In the iso-singlet case ($a=0$) this quantity is nonzero.
If the $s$-quark mass is not degenerate with the masses of $u-$ and $d$-quarks
then $G^{(1)}_{8P}$ is also nonzero. 
For $a=1,\dots,7$ the constants 
are identicaly equal to zero.

Calculation of  $G^{(1)}_{aP}$ is quite simple. 
The result of the action of the vertex operator $F_{00}(x|B)$
onto the quark propagator can be found as follows
(in our case $\xi_f=1/2$ in Eq.~(\ref{xi})):
\begin{eqnarray}
\label{vertprop}
F_{00}(x|B)S(x,x|B)&=&
\int\limits_0^1dt 
\exp\left\{
\frac{t}{4\Lambda^2}
\left(\stackrel{\leftarrow}{\nabla}_x-
\stackrel{\rightarrow}{\nabla}_y
\right)^2
\right\}S(y,x|B)|_{x=y}
\nonumber\\
&=&\int\limits_0^1dt \int\frac{d^4p}{(2\pi)^4}\tilde H(p)
e^{-tp^2/\Lambda^2}.
\end{eqnarray}
Furthermore,
\begin{eqnarray}
\label{trg5prop}
{\rm Tr}i\gamma_5\tilde H(p)=
\mp\frac{2im}{v\Lambda^2}\int\limits_0^1ds
e^{-sp^2/2v\Lambda^2}\left(\frac{1-s}{1+s}\right)^{m^2/4v\Lambda^2}
\frac{s^2}{1-s^2}.
\end{eqnarray}
Here the sign $``\mp''$ relates to self- and anti-self-dual configurations
of the vacuum field.

Substitution of  Eqs.~(\ref{trg5prop}) and (\ref{vertprop}) in
Eq.~(\ref{j08}) and integration over the loop momentum $p$ and the
parameter $t$ leads to the following integral representation
\begin{eqnarray}
\label{j08-1}
G^{(1)}_{aP}=\pm i \frac{\Lambda^3}{2\pi^2} \sum_{f}\lambda^a_{ff}R_f, \ \ 
R_f=
\sum_{v}v\frac{m_f}{\Lambda}
\int\limits_0^1\frac{dss}{(2v+s)(1-s^2)}\left({1-s\over 1+s }\right)^{m_f^2/4v\Lambda^2}, 
\end{eqnarray}
where summation over $f$ and $v$ corresponds to traces in the flavour
and colour spaces. As indicated in Eq.~(\ref{b-eta}), 
$v$ is a diagonal matrix 
relating to the direction of the vacuum field in colour space.

Eq.~(\ref{j08-1}) results in the following momentum representation of  
the two-point correlation function, Eq.~(\ref{gam2})  
\begin{eqnarray*}
&&\Lambda^{-2}\tilde\Gamma^{(2)}_{aP,bP}(p^2)=
\tilde\Pi_{ab}(p^2)=\Pi_{ab}(p^2)
-\delta\Pi_{ab}(p^2),
\nonumber
\end{eqnarray*} 
where 
\begin{eqnarray}
\label{deltaP}
&&\delta\Pi_{ab}(p^2)=-\frac{8}{3\pi^4}(\Lambda R)^4T^{ab}F(p^2),
\ \ 
T^{ab}=\sum\limits_{ff'}\lambda^a_{ff}\lambda^b_{f'f'}R_fR_{f'},
\nonumber\\
&&F(p^2)=\int\limits_0^1dt\sqrt{1-t^2}\int\limits_0^1dss
\cos\left(\sqrt{4p^2R^2t^2s}\right)
\left[\frac{3\pi}{2}-3\arcsin\sqrt{s}-(5-2s)\sqrt{s(1-s)}\right],
\nonumber\\
&&T^{00}=\frac{1}{3}(2R_u+R_s)^2,
\ 
T^{88}=\frac{2}{3}(R_u-R_s)^2,
\ 
T^{08}=\frac{\sqrt{2}}{3}(2R_u+R_s)(R_u-R_s),
\nonumber\\
&&\Pi_{00}=\frac{1}{3}(2\Pi_u+\Pi_s),
\
\tilde\Pi_{88}=\frac{1}{3}(\Pi_u+2\Pi_s),
\
\Pi_{08}=\frac{\sqrt{2}}{3}(\Pi_u-\Pi_s).
\end{eqnarray}
The function $\Pi_{f}(p^2)$ has been calculated 
in Ref.~\cite{PRD96},
\begin{eqnarray}
\label{polx1}
\Pi_{f}(p^2)=
&-&\frac{1}{4\pi^2}{\rm Tr}_v
\int_0^1dt_1\int_0^1dt_2\int_0^1ds_1\int_0^1ds_2
\left(\frac{1-s_1}{1+s_1}\right)^{\frac{m_f^2}{4v\Lambda^2}}
\left(\frac{1-s_2}{1+s_2}\right)^{\frac{m_{f}^2}{4v\Lambda^2}}
\nonumber\\
&\times&\left[-\frac{p^2}{\Lambda^2}
\frac{F_1(t_1,t_2,s_1,s_2)}{\varphi_2^4(t_1,t_2,s_1,s_2)}+
\frac{m^2_f}{\Lambda^2}
\frac{F_2(s_1,s_2)}{(1-s_1^2)(1-s_2^2)
\varphi_2^2(t_1,t_2,s_1,s_2)}\right.
\nonumber\\
&+&\left.\frac{2v(1-4v^2t_1t_2)F_3(s_1,s_2)}
{\varphi_2^3(t_1,t_2,s_1,s_2)}\right]
\exp\left\{-\frac{p^2}{2v\Lambda^2}\varphi(t_1,t_2,s_1,s_2)\right\},
\nonumber\\
\varphi&=&\frac{\varphi_1(t_1,t_2,s_1,s_2)}{\varphi_2(t_1,t_2,s_1,s_2)},
\
\varphi_1=\frac{1}{2}v(t_1+t_2)[s_1+s_2]+s_1s_2,
\nonumber\\
\varphi_2&=&2v(t_1+t_2)(1+s_1s_2)+(1+4v^2t_1t_2)(s_1+s_2),
\nonumber\\
F_1&=&(1+s_1s_2)[(s_1+v(t_1+t_2))(s_2+v(t_1+t_2))+v^2(t_1-t_2)^2s_1s_2],
\nonumber\\
F_2&=&(1+s_1s_2)^2, 
\
F_3=2(1+s_1s_2). \nonumber
\end{eqnarray}

The masses of  $\eta$ and $\eta'$ can be calculated by means of 
Eqs.~(\ref{mass-eq}), which take the form 
\begin{eqnarray}
\label{meta}
&&1+\frac{g^2}{9} \Pi_{\eta}(-M^2_{\eta})=0,
\
1+\frac{g^2}{9}\Pi_{\eta'}(-M^2_{\eta'})=0,
\nonumber\\
&&\Pi_\eta=\frac{1}{2}
\left[\tilde\Pi_{00}+\tilde\Pi_{88}
-\sqrt{(\tilde\Pi_{00}-\tilde\Pi_{88})^2
+4\tilde\Pi_{08}^2}
\right],
\nonumber\\
&&\tilde\Pi_{\eta'}=\frac{1}{2}
\left[\tilde\Pi_{00}+\tilde\Pi_{88}
+\sqrt{(\tilde\Pi_{00}-\tilde\Pi_{88})^2
+4\tilde\Pi_{08}^2}
\right].
\nonumber
\end{eqnarray}
where $\tilde \Pi=\Pi-\delta \Pi$ includes both diagrams in Fig.\ref{fig:diagr}.
These equations can be written more transparently if we introduce the 
mixing angle
\begin{eqnarray}
\label{mixing}
&&\eta=\eta_8\cos\theta+\eta_0\sin\theta,
\nonumber\\
&&\eta'=\eta_0\cos\theta-\eta_8\sin\theta,
\end{eqnarray}
where the angle $\theta$ is a function of momentum,
\begin{eqnarray}
\label{theta}
&&\tan 2\theta(p^2)=2\tilde\Pi_{08}(p^2)/
\left(\tilde\Pi_{88}(p^2)-\tilde\Pi_{00}(p^2)\right).
\end{eqnarray}
The polarization functions of $\eta$ and $\eta'$ and their derivatives with
respect to $p^2$ take the form
\begin{eqnarray}
\label{polee'}
&&\Pi_\eta=
\tilde\Pi_{88}\cos^2\theta + 
\tilde\Pi_{00}\sin^2\theta+
\tilde\Pi_{08}\sin 2\theta,
\nonumber\\
&&\Pi_{\eta'}=
\tilde\Pi_{00}\cos^2\theta +
\tilde\Pi_{88}\sin^2\theta -
\tilde\Pi_{08}\sin 2\theta,
\nonumber\\
&&\Pi'_\eta=
\tilde\Pi'_{88}\cos^2\theta + 
\tilde\Pi'_{00}\sin^2\theta+
\tilde\Pi'_{08}\sin 2\theta,
\nonumber\\
&&\Pi'_{\eta'}=
\tilde\Pi'_{00}\cos^2\theta +
\tilde\Pi'_{88}\sin^2\theta -
\tilde\Pi'_{08}\sin 2\theta,
\end{eqnarray}
where for computing derivatives Eq.~(\ref{theta}) has been used.

With the values of the parameters
\begin{eqnarray}
\label{parr1}
m_u=m_d=177.85 \ {\rm MeV}, 
\
m_s=400.98 \ {\rm MeV},
\
\sqrt{B}=469.52 \ {\rm MeV},
\
g=8.94,
\end{eqnarray}
fitted from the masses of $\pi$, $\rho$, $K$ and $K^\star$ mesons
such that 
\begin{eqnarray}
\label{mpv}
&&M_\pi=140{\rm MeV},  M_K=496 {\rm MeV}, f_\pi=129.9 {\rm MeV}, f_K=150.8 {\rm MeV},
\nonumber\\
&& M_\rho=M_\omega=770 {\rm MeV} , M_{K^*}=890 {\rm MeV}, M_\phi= 1035 {\rm MeV}, 
\end{eqnarray}
we find that
\begin{eqnarray}
&&M_\eta=640 \ {\rm MeV},  \ M_{\eta'}=950 \ {\rm MeV},
\nonumber\\ 
&& h_{\eta}=4.72, \ h_{\eta'}=2.55
\nonumber
\end{eqnarray}
for
\begin{eqnarray}
\sqrt{B}R=1.56.
\label{sbr}
\end{eqnarray}
The parameter $R$ was fitted to $\eta'$ mass. The domain model value for this dimensionless parameter
quoted in the first part of the paper  $\sqrt{B}R=1.24$
is close to 
 Eq.(\ref{sbr}).
Thus we see that all features of the spectrum of light vector and 
pseudoscalar mesons 
usually associated with chiral symmetries are correctly reproduced 
by the model quantitatively. The origin of the splitting between 
the pseudoscalar and vector mesons was discussed in detail in \cite{PRD96}.
The new feature is the shift in the masses of $\eta'$ and $\eta$ with 
respect to the other pseudoscalar
mesons and their mutual splitting, which occurs here due to the additional 
contribution  of Eq.~(\ref{deltaP}) to their polarization functions.
The momentum dependent mixing angle
takes different values at the scale of $\eta$ and $\eta'$ masses
\begin{eqnarray}
\label{mixang}
\theta_\eta=\theta(-M^2_\eta)=-19.8^\circ, \  
\theta_{\eta'}=\theta(-M^2_{\eta'})=46.1^\circ.
\end{eqnarray}
It is notable that  $\theta_\eta$
coincides with the value of the mixing angle in the naive quark model,
while $\theta_{\eta'}$ is completely different from $\theta_\eta$
both in sign and magnitude.

We quote also the result for the weak decay constants
of $\eta$ and $\eta'$: 
\begin{eqnarray*}
&&f^0_\eta=4.13 \  {\rm MeV}, \  f^8_\eta=165.7 \ {\rm MeV}, 
\ f^\theta_\eta=154.5 \ {\rm MeV}, 
\nonumber\\
&&f^0_{\eta'}=288.6 \ {\rm MeV}, \  f^8_{\eta'}=23.67 \ {\rm MeV}, 
\ f^\theta_{\eta'}=183.3 \ {\rm MeV}.
\nonumber 
\end{eqnarray*}
These constants are defined by the matrix elements
\begin{eqnarray*}
&&\langle0|J^a_{5\mu}(0)|\phi({\bf p})\rangle = ip_\mu f^a_\phi/\sqrt{2},
\nonumber\\
&&J^a_{5\mu}=\bar q \gamma_\mu\gamma_5 t^a q, \ t^0=\lambda^0/2,
 \ t^8=\lambda^8/2,
\nonumber
\end{eqnarray*}
and $f^\theta_\phi$ corresponds to the mixings:
\begin{eqnarray*}
&&f^\theta_\eta=f_\eta^8\cos\theta +f_\eta^0\sin\theta,
\nonumber\\ 
&&f^\theta_{\eta'}=f_\eta^0\cos\theta -f_\eta^8\sin\theta.
\nonumber
\end{eqnarray*} 
Here
\begin{eqnarray*}
&&f^a_\eta=h_\eta\left[f_8^a\cos\theta +f_0^a\sin\theta\right],
\ 
f^a_{\eta'}=h_{\eta'}\left[f_0^a\cos\theta -f_8^a\sin\theta\right],
\nonumber\\
&&f_0^0=\frac{1}{3}(2f_u+f_s), \ f_8^8=\frac{1}{3}(f_u+2f_s),
 \ f_8^8=\frac{\sqrt{2}}{3}(f_u-f_s).
\end{eqnarray*} 
The function $f_f(-M^2)$
was calculated in \cite{PRD96},
\begin{eqnarray*}
&&f_f=\frac{m_f}{4\pi^2}\sum\limits_v
\int\limits_0^1\!\int\limits_0^1\!\int\limits_0^1
\frac{dt ds_1 ds_2 (1+s_1s_2)}{[s_1+s_2+2vt(1+s_1s_2)]^3}
\left[\frac{(1-s_1)(1-s_2)}{(1+s_1)(1+s_2)}\right]^{m_f^2/4v\lambda^2}
\nonumber\\
&&\ \ \ \ \ \ \ \ \ \ \ \ \ \ \ \         
\times\left[2vt +\frac{s_1}{1-s_1^2}+\frac{s_2}{1-s_2^2}\right]
\exp\left(\frac{M^2}{2v\Lambda^2}\Psi(t,s_1,s_2)\right),
\nonumber\\
&&\Psi=\frac{s_1s_2+vt(s_1+s_2)/2}{s_1+s_2+2vt(1+s_1s_2)}.
\end{eqnarray*}
As we have already mentioned, the quark masses here should be considered as 
constituent quark masses. The massless limit is ill-defined due to the 
contribution of zero modes
to the quark propagator. However, as is discussed in \cite{PRD96}, a  
peculiar feature is that all the necessary shifts and splittings in the 
meson spectrum occur explicitly due to the strong (and, 
for pseudoscalar, scalar,vector and axial vector states, very different) 
dependence of meson masses on the quark masses, which is driven entirely  
by the zero modes.  In order to visualize  
this picture in a quantitative manner, including now also 
the $\eta$ and $\eta'$, 
let us  consider a regime in the model which  
can be called  ``a chiral limit".

\subsection{Massless pseudoscalar octet.}

This limit can be defined in terms of composite meson fields
as the condition 
that the $\pi$ and $K$ mesons become massless 
particles. Within the above-formulated meson 
theory this requirement is fulfilled if constituent quark masses
satisfy the relation
\begin{eqnarray*}
&&m_s=m_d=m_u=m^*
\nonumber
\end{eqnarray*}
and the mass $m^*$ is defined from the equation for pion mass with $m_\pi=0$
substituted, that is
\begin{eqnarray}
&&1+\frac{g^2}{9\Lambda^2}\Pi_\pi(0)=0.
\label{pion0}
\end{eqnarray}
Since in this limit flavour $SU(3)$ becomes exact symmetry  
the kaons are also massless.
Furthermore, as follows from Eqs.~(\ref{deltaP}) 
\begin{eqnarray*}
\Pi_{08}=T^{08}=T^{88}\equiv 0,
\end{eqnarray*}
which  means that the $\eta$ meson is degenerate with the $\pi$ and $K$.
The mixing angle between $\eta_0$ and $\eta_8$
is equal to zero and the $\eta'$ meson corresponds to 
a pure flavour singlet state.
As seen from Eqs.~(\ref{deltaP}),
\begin{eqnarray*}
\delta\Pi_{00}\propto T^{00}F(p^2)\not=0
\nonumber
\end{eqnarray*} 
and the mass of the singlet state is split from the flavour octet states.
Vector mesons are massive in this regime.
Numerically one finds
\begin{eqnarray*}
&&m^*=168.83 \ {\rm MeV}, 
\nonumber\\ 
&&M_\eta=M_\pi=M_K=0, \ M_{\eta'}=897 \ {\rm MeV}, 
\nonumber\\
&&h_\eta=h_\pi=h_K=4.28, \ h_{\eta'}=.63, 
\nonumber\\
&&f_\eta^8=f_K=f_\pi=127.76 \ {\rm MeV}, \ f^0_{\eta'}=281.87 \ {\rm MeV},
\nonumber\\
&&f^0_\eta=f^8_{\eta'}=0.
\nonumber
\end{eqnarray*}  
and 
\begin{eqnarray*}
M_\rho=M_{K^*}=M_\omega=762 {\rm MeV}, \ \ M_\phi=950 {\rm MeV}.
\end{eqnarray*}
This result displays a typical chiral limit picture with both 
flavour $SU_{\rm R}(3)\times SU_{\rm L}(3)$ and $U_A(1)$ symmetries
correctly implemented {\it de facto}:
while the masses of octet states are drastically reduced from physical values 
to zero their  decay constants are subject  only to minor change.
Simultaneously we see that the $\eta'$ does not look like a 
"Goldstone" particle:
both its mass and decay constant $f^0_{\eta'}$ are practically unchanged.
The vector states are subject to minor change also.
For completeness we mention that the scalar and axial vector particles as  
ground state mesons are absent in the spectrum in this model, but they appear 
in the hyperfine structure of orbital excitations of vector mesons with 
quantitatively correct masses~\cite{PRD96},
which stays unchanged also in the chiral limit described here.

In this picture $m^*$ looks like the condensate part of the constituent
quark mass. This impression can be enhanced by 
estimating  the current quark masses and their ratio,
\begin{eqnarray}
&&\mu_u=m_u-m^*=9.02 \ {\rm MeV}, \ \mu_s=m_s-m^*=232.15 \ {\rm MeV},
\nonumber\\ 
&&\mu_s/\mu_u=25.73.
\nonumber
\end{eqnarray}
The ratio, which is the only meaningful quantity,
practically coincides with the generally accepted value $\mu_s/\mu_u=25$.

\section{CONCLUSIONS AND DISCUSSION}

We have shown that the correlation between the
chirality of low-lying Dirac modes and the
duality of the domain-like background field indeed
drives the spontaneous breakdown of flavour chiral
symmetry, as was hinted at in \cite{NK2002}, and
as indicated in lattice calculations \cite{Xstudies}.
We have also shown that the mechanism generating
the $\eta-\eta'$ mass splitting is the same as that
causing area law confinement in the domain model, namely
the finite range correlations induced by the domain mean field.
In more detail, 
we have extracted the parity-odd part of the 
logarithm of the quark determinant and seen that
the axial anomaly is recovered. The chirality properties
of the Dirac modes in domains generate
the anomaly in this context. We see however that 
in this formulation with domains with bag-like
boundary conditions both the symmetric zeta and the
asymmetric eta functions are necessary to obtain this
result. We have then explored the consequences of the
anomaly for the realisation of chiral symmetry in the
domain model. The contribution of the anomaly to the
free energy of an ensemble of domains leads to
a spontaneous breaking of $Z(2)$ symmetry rather than
a $U_A(1)$ for one massless fermion flavour. 
Quark condensation occurs because of the spectral
asymmetry coming from the bag-like boundary
conditions. In the multiflavour case, the Abelian
nature of the axial anomaly guarantees that 
the discrete $Z(2)$ symmetry remains spontaneously
broken in addition to the correct continuous non-singlet
$SU(N_f)_L\times SU(N_f)_R$ symmetries. 
One would thus naively expect only $N_f^2-1$ Goldstone bosons
based on the number of continuous degenerate directions
in the space of ground states of an ensemble of domains.

On this basis we expect massless pions, kaons and eta mesons
in the chiral limit. The domain model manifests correlations 
Fig.\ref{fig:diagr}b in the singlet channel
which generates the splitting in $\eta-\eta'$ masses via
the Witten mechanism. 
The contribution, Fig.\ref{fig:diagr}b,
is entirely driven by the correlators of the background field 
Eq.~(\ref{corfunct}).
Simultaneously the background field correlators are also entirely responsible 
for the area law~\cite{NK2001}. Thus the origin of both mechanisms is 
identical. In the $N_{\rm c}\to\infty$ limit the $\eta'$ is 
massless since the contribution of Fig.\ref{fig:diagr}b is $1/N_{\rm c}$
suppressed with respect to  Fig.\ref{fig:diagr}a.

We have seen that the calculational scheme does work well phenomenologically, 
but the dimensionless parameter turns out to be $BR^2\approx O(1)$ such that 
for a self-consistent consideration of meson physics one
needs to use the propagators of the domain background field: 
neglecting boundary conditions
inside quark loops is not consistent with $BR^2\approx O(1)$. 
There is no clear separation of two scales as has already been observed in the 
consideration of static parameters of the vacuum --  quark and gluon 
condensates, the string constant and the topological susceptibility
within the domain model.

Nevertheless, the calculations presented in the final section resemble 
two features of the domain model we have addressed in the first part of 
the article. The first feature is quite obvious:
in both cases the additional contributions to the correlators
(polarization functions)  for $\eta'$  and, if  $m_u\not=m_s$ also for $\eta$, 
are crucial for their splitting from the other pseudoscalar mesons, 
thus resolving this aspect of the $U_A(1)$ problem.
Certainly, the successful (but not self-consistent) quantitative 
description provided by 
the purely homogeneous background field need not {\it ad hoc} be equally 
successful in the domain model, and verification 
of this is one of our first priorities. 
The second feature is not quite so obvious.  The
splitting between pseudoscalar and vector meson
masses  in the case of the purely homogeneous field is determined by the
singular behaviour of the quark condensate for $m\to 0$
(as mentioned, the 
condensate diverges in the massless limit due to a 
continuum of zero modes~\cite{Smilga,EfNed}).
This singular behaviour is not present in the domain model -- 
zero modes do not exist at all and the limit $m\to0$
is regular.
However a quark condensate is generated in the domain model by the 
asymmetry in the spectrum of the Dirac operator, and this condensate diverges  
for $R\to\infty$ as follows from 
Eqs.~(\ref{cond0}) and (\ref{cond-f})
and as shown in Fig.~\ref{fig:cond}, as discussed above. 
This divergence is expected to  
play the same role for correlators
of the domain model as  $m\to 0$ in the homogeneous field, 
thus generating  strong pseudoscalar-vector 
splitting.  The value of quark condensate given in Eq.~(\ref{vcond-f})
corresponds to $R/R_0=1$ in Fig.~\ref{fig:cond}.
But, unlike the model based on a purely homogeneous field,
the domain model manifests spontaneous breaking
of flavour chiral symmetry and hence has the potential to reproduce a
genuine picture of the chiral limit.
This is only mimicked in the 
homogeneous field considerations above, though in a surprisingly 
successful manner.

The zeta function calculation nonetheless remains
incomplete, where the contributions that should eliminate
the chiral angle dependent parity even part of the
quark determinant's logarithm need to be calculated. The observation
of lack of separation of scales, mentioned above, also means that a more
careful study of collective modes within the domain model
{\it per se} is necessary to put the phenomenological results above on 
a sound footing. In the same context, explicit correlation functions
in the domain model would enable a study of the 
(anomalous) Ward identities and issues \cite{ua1papers}
related to the realisation of Goldstone's theorem 
(or otherwise) in the $U_A(1)$ sector. 

There are two issues naturally related to this work and requiring at least some
preliminary comment within the constraints of the calculations realised
thus far.

At first superficial glance, the mechanism of quark condensate 
generation through the asymmetry in the
spectrum of the Dirac operator looks to be different to the approach of
Banks and Casher \cite{BC80} where the spectrum is
 symmetric and the condensate arises due to a finite density of
chiral zero modes. However, these seemingly different formulations in fact
have more in common.
In particular, in the thermodynamic limit the system is characterised by a
finite density of low-lying (but non-zero)
modes which are strongly chiral. Even if for a given fixed domain the
spectrum is asymmetric, as indicated
by the spectral asymmetry function $\eta$,  
on average all positive and negative non-zero
eigenvalues appear in the ensemble in a symmetric way, and
neither left nor right chirality modes prevail in the ensemble. 
This can be seen in the symmetric distribution of the local chirality 
parameter given in our previous work \cite{NK2002}.
The chiral condensate arises due to the existence of a finite density of 
strongly chiral eigenmodes both in the Banks-Casher formulation and 
in the model under consideration.
The difference is that, in the domain model
the corresponding eigenvalues are nonzero in the thermodynamic limit
and none of the fermionic modes are purely chiral,
rather they can be characterised by their average chirality
with a definite sign correlated with the duality of the mean gluonic field in
the domain: the lower the Dirac operator eigenmode the closer  this
average chirality to $\pm1$.
Thus, we would take the liberty to say that the domain model gives a
"smeared" realisation of the Banks-Casher scenario.

Another interesting question inspired by the chiral boundary condition
used in the model is the manifestation or otherwise of
features seen in chiral bag models of the nucleon.
How much is in common here beyond the similar boundary condition?
There is no simple answer to this question yet.  We recall
that the domain model is formulated in Euclidean four-dimensional
space, colourless hadrons (if any) are not associated with domains 
themselves but are anticipated to arise as collective excitations 
of quantum fluctuations of quark (and possibly gluon)
fields in the domain ensemble.
The description of colourless hadrons requires analytical continuation to 
physical Minkowski space.
Quark and gluon fluctuations in this approach are localised both in space 
and time: no asymptotic particle-like states can be associated with them. 
The model requires substantial use of methods of nonlocal
quantum field theory. Various issues related to quantisation, unitarity, 
causality, Froissart-type bounds at high energies
and interpetation of nonlocal fields as fluctuations localised in space
and time can be found in~\cite{Efimov,compcond,EfGanb,Smekal,Fainberg}.
The nonlocality appears here primarily due to the
presence of strong background gluon fields which eliminate the pole in the 
momentum space quark propagator ~\cite{Leutw80,PRD95,PRD96} rendering it
an entire analytical function.
The nature of coloured fluctuations as localised in space and time
prohibits a straightforward  resolving in the domain model of the manifestation
of such interesting features as baryon number
fractionalisation in chiral bags \cite{NiSe86}.
This phenomenon appears due to
asymmetry (and a corresponding $\eta$ invariant)
in the spectrum of the Dirac Hamiltonian, rather than the Euclidean Dirac
operator, after a rearrangement of energy levels in the solitonic background
chiral bag. There is no automatic one-to-one
correspondence between phenomena in the domain model and those of the
chiral bag model for the nucleon.
In particular, the answer to the question of baryon number is
intimately connected to the analytical properties of
propagators (for example see \cite{SSS98}) and thus to the mode of dynamical confinement
and the realisation of hadrons as propagating excitations in the
domain ensemble, the subject of further work.

\section*{ACKNOWLEDGEMENTS}
S.N.N. was supported by the DFG under contract SM70/1-1 and, 
partially, by  RFBR grant 01-02-1720. 
A.C.K. is supported by a grant of the Australian Research Council.
We acknowledge the hospitality of the Institute
for Theoretical Physics III, University of Erlangen-Nuremberg
where substantial parts of this work was done.
In particular we acknowledge fruitful discussions with
Frieder Lenz, Michael Thies, Lorenz von Smekal, Jan Pawlowski
and Andreas Schreiber. We also thank G.V. Efimov for numerous discussions and
explanations, K. Kirsten, G. Dunne and S. Bilson-Thompson.

\appendix
\section{Expansion of Kummer functions}

\subsection{Asymptotic form of Kummer function 
$M(k^2x^2/(4z),k+1,z)$ for $k\gg 1$, $z, \ x$ -- fixed.}

We use the representation~\cite{magnus}
\begin{eqnarray*} 
M(a/4z,k+1,z)=\frac{k!}{\Gamma(a/4z)}\int_0^\infty dt t^{a/4z-1}(zt)^{-k/2}
e^{-t}I_k(2\sqrt{zt})
\end{eqnarray*}
and change the integration variable $ 2\sqrt{zt}=s$
giving
\begin{eqnarray*} 
\label{eq3}
M(a/4z,k+1,z)=\frac{k!z^{-a/4z}2^{k+1-a/2z}}{\Gamma(a/4z)}
\int_0^\infty ds s^{a/2z-k-1}e^{-s^2/4z}I_k(s).
\end{eqnarray*}
The asymptotic behaviour of 
\begin{eqnarray*} 
{\cal M}_k=\int_0^\infty ds s^{a/2z-k-1}e^{-s^2/4z}I_k(s)
\end{eqnarray*}
at $a\propto k^2$, $k\gg1$ can be found by the saddle point method.
Denoting $\alpha=a/2z$, $\zeta=2z$ and 
$f(s)=[1-(k+1)/\alpha]\ln(s)-s^2/2\alpha\zeta$,
we arrive at 
\begin{eqnarray*} 
{\cal M}_k=\int_0^\infty ds e^{\alpha f(s)}I_k(s)
\end{eqnarray*}
with $f(s)$  having a maximum at  
\begin{eqnarray*} 
s_0=\sqrt{\zeta(\alpha-k-1)}.
\end{eqnarray*}
Expanding the exponent about the saddle point and using the addition theorem 
\begin{eqnarray*}
\label{eq17}
&&\left(\frac{z_1-z_2e^{-i\phi}}{z_1-z_2e^{i\phi}}\right)^\nu I_\nu(\omega)=
\sum_{n=-\infty}^{\infty}(-1)^n I_n(z_2)I_{\nu+n}(z_1)e^{in\phi},\\
&&\omega=\sqrt{z_1^2+z_2^2-2z_1z_2\cos(\phi)}
\end{eqnarray*}
with $\phi=0$ and $\pi$ one obtains
the following representation 
\begin{eqnarray} 
M(a/z,k+1,z)&=&\sqrt{2\pi z}\frac{k!z^{-a/4z}2^{k+1-a/2z}}{\Gamma(a/4z)}
\left[a-2z(k+1)\right]^{a/4z-(k+1)/2}e^{-a/4z+(k+1)/2}e^{z/4}
\nonumber\\
&\times&\sum_{n=-\infty}^{\infty}I_{2n+k}(s_0)I_n(z/4)
\left[
1+W_1(x,z,n)/k+W_2(x,z,n)/k^2
\right]
\nonumber\\
&&
+
\sum_{n=-\infty}^{\infty} I_{2n+k+1}(s_0)I_{n}(z/4)\frac{W_3(x,z)}{k},
\\
&&W_1=\frac{z(z-2n+3)}{6x^2}
\nonumber\\
&&W_2= \frac{z}{72x^4} \left[(27 z-8 z n+z^3+15 z^2+4 z n^2) 
\sqrt{1+x^2}+8 z n^2 x^2-12 x^2
\right.
\nonumber\\
&&\left.
+15 z^2-9 z x^2+2 z^2 n x^2+8 n^3 x^2+z^3 x^2+4 z^2 x^2-56 z n+27 z
\right.
\nonumber\\
&&\left.
-48 n^2 x^2+46 n x^2+26 z n x^2+4 z n^2+z^3\right]
\nonumber\\
&&W_3=\frac{z(z+3)}{6x}
\nonumber\\
s_0&=&\sqrt{a-2z(k+1)}, \ \ a=k^2x^2, \ \ x=t\rho,
\nonumber
\label{Masymprelim}
\end{eqnarray}
and calculation of the asymptotic form of $M$ is reduced to
the asymptotic decomposition of
$I_{2n+k}(s_0)$ and $I_{2n+k+1}(s_0)$ for $k\gg n$.

Decomposition of the factors in front of the sum in
Eq.(\ref{Masymprelim}) at $k\gg1$ gives
\begin{eqnarray*}
&&\frac{k!z^{-a/4z}2^{k+1-a/2z}}{\Gamma(a/4z)}
\left[a-2z(k+1)\right]^{a/4z-(k+1)/2}e^{-a/4z+(k+1)/2}e^{z/4}\sqrt{2\pi z}=
\nonumber\\
&&\frac{k!2^k}{(kx)^k}e^{z/4+z/2x^2}\left[1+\frac{z(3x^2+z)}{3x^4k}+
\frac{z(3x^6+27x^4z +12x^2z^2 +z^3)}{18x^8k^2}+O(1/k^3)\right]
\end{eqnarray*} 
The representation
\begin{eqnarray*} 
I_{\nu}(s_0)=\frac{1}{\sqrt{\pi}}
\frac{s_0^\nu}{2^\nu\Gamma(\nu+1/2)}\int_{-1}^1 dt e^{-s_0t}(1-t^2)^{\nu-1/2}
\end{eqnarray*}
is now suitable for determining an asymptotic decomposition of the
$I_{2n+k}(s_0)$ again by means of the saddle point approximation. 
With 
\begin{eqnarray*} 
g_1(x,n,z)& =&
 -\frac{8z^2+4x^2z(2+4n+z)+x^4(-1+16n^2+8z)}{8x^4\sqrt{1+x^2}},
\\
g_2(x,n,z) &= &
\frac{ [8 z^2+4 x^2 z (2+4 n+z)+x^4 (-1+16 n^2+8 z)]^2}{128x^8(1+x^2)}
\nonumber\\
&-&\frac{1}{24(1+x^2)^{3/2}x^6} 
\left\{32 z^3+48 x^2 z^2 (1+n+z)+x^6 [-1+6 n-32 n^3-3 z+48 n (1+n) z+24 z^2]
\right.
\nonumber\\
&+&\left.12 x^4 z [z (6+z)+n (4+6 z)]\right\}.
\end{eqnarray*}
and
\begin{eqnarray*}
&&h_1(x,n,z) = \frac{2(x^2-2)\sqrt{1+x^2}-5}{24(1+x^2)^{3/2}}
\\
&&h_2(x,n,z)= \frac{1}{1152(1+x^2)^3}\left[-8-68 x^2-720 z+480 n-960 x^2 n) 
\sqrt{1+x^2}
\right.
\\
&&\left. +17-144 x^2+276 x^4+52 x^6+384 n-576 x^2 n
-1152 x^4 n-192 x^6 n-576 z x^2-576 z\right]
\end{eqnarray*} 
we obtain the following full expression 
for the asymptotics of the Kummer function up to $O(1/k^2)$
\begin{eqnarray*} 
M(a/4z,k+1,z)&=&
\frac{k!2^k}{(kx)^k}\frac{1}{\sqrt{2\pi k}}\frac{1}{(1+x^2)^{1/4}}
\exp\left\{  k\sqrt{x^2+1}+k\ln\left(\frac{x}{1+\sqrt{x^2+1}}\right)
\right\}  
\nonumber\\ 
&\times&\exp\left\{  z/4+z/2x^2-z\frac{\sqrt{x^2+1}}{x^2}\right\}
\left[1+\frac{z(3x^2+z)}{3x^4k}+
\frac{z(3x^6+27x^4z +12x^2z^2 +z^3)}{18x^8k^2}+O(1/k^3)\right] 
\nonumber\\
&\times&
\sum_{n=-\infty}^\infty I_n(z/4)
 \left(\frac{x}{1+\sqrt{x^2+1}}\right)^{2n}
\nonumber
\\
&\times & \left\{\left[ 1+\frac{2z+1-4n}{4(1+x^2)k}+
 \frac{16n^2(3-2x^2)+8n(2x^2-3-10z)+16zx^2+36z-2x^2+20z^2+3}{32(1+x^2)^2k^2} 
\right] \right.
\nonumber\\
&~&\times \left. \left[1+\left(\frac{1}{24}-2n^2\right)\frac{1}{k}
+\left(\frac{1}{1152}-\frac{n}{12}-\frac{n^2}{12}+\frac{4n^3}{3}+2n^4\right)
\frac{1}{k^2}\right]\right.
\nonumber
\\
&~&\times \left.
\left[
1+\frac{16n^2-1}{8k}+\frac{768n^4-512n^3-96n^2+96n-13}{384k^2}
\right]
\right.
\nonumber
\\
&~&\times \left.
\left[ 1+\frac{g_1(x,n,z)}{k}+ \frac{g_2(x,n,z)}{k^2}\right]
\left[ 1+\frac{h_1(x,n,z)}{k}+ \frac{h_2(x,n,z)}{k^2}\right]
\right.
\nonumber
\\&~& \left.\times\left[ 1+\frac{W_1(x,n,z)}{k}+ \frac{W_2(x,n,z)}{k^2}\right]
\right.
\nonumber
\\&~& \left.+\frac{z(z+3)}{6k(1+\sqrt{1+x^2})}
\left[ 1+\frac{2z-1-4n}{4(1+x^2)k}
 \right]
\left[1+\left(\frac{1}{24}-2(n+1/2)^2\right)\frac{1}{k}
\right]
\left[
1+\frac{16(n+1/2)^2-1}{8k}
\right]
\right.
\nonumber
\\
&~&\times \left.
\left[ 1+\frac{g_1(x,n+1/2,z)}{k}\right]
\left[ 1+\frac{h_1(x,n+1/2,z)}{k}\right]
+O(1/k^3)
\right\}
\end{eqnarray*}
The series over $n$ can now be resummed in terms of  
the generating function of Bessel functions and its derivatives,
\begin{eqnarray*}
&&\sum_{n=-\infty}^\infty v^n  I_n(u)=\exp\left\{ \frac{u}{2}
\left( v+\frac{1}{v}\right)\right\},
\nonumber\\
&&\sum_{n=-\infty}^\infty n^lv^nI_n(u)=\left[v\frac{d}{dv}\right]^l\exp
\left\{ \frac{u}{2}\left( v+\frac{1}{v}\right)\right\}
\end{eqnarray*} 
where
\begin{eqnarray*}
u=z/4, \ \ v=\left(\frac{x}{1+\sqrt{x^2+1}}\right)^{2},  \ \ \ l=0,1,2,3,4.
\end{eqnarray*} 
Moreover, the presence of the factor
$$\exp\left\{  z/4+z/2x^2-z\frac{\sqrt{x^2+1}}{x^2}\right\}$$
allows further simplification
\begin{eqnarray*}
\exp\left\{  z/4+z/2x^2-z\frac{\sqrt{x^2+1}}{x^2}\right\}
\sum_{n=-\infty}^\infty \left(\frac{x}{1+\sqrt{x^2+1}}\right)^{2n}  I_n(z/4)
&=&\exp\left\{\frac{z}{2}\frac{\sqrt{x^2+1}-1}{\sqrt{x^2+1}+1} \right\}.
\end{eqnarray*} 
The remaining formulae required can be obtained by differentiation.
The final result is then
\begin{eqnarray*} 
M(a/4z,k+1,z)&=&
\frac{k!2^k}{(kx)^k}\frac{1}{\sqrt{2\pi k}}\frac{1}{(1+x^2)^{1/4}}
\exp\left\{  k\sqrt{x^2+1}+k\ln\left(\frac{x}{1+\sqrt{x^2+1}}\right)
\right\}  
\nonumber\\ 
&\times&\exp\left\{ \frac{z}{2}\frac{\sqrt{x^2+1}-1}{\sqrt{x^2+1}+1}\right\} 
\left[1+\frac{M_1(x,z)}{k}+\frac{M_2(x,z)}{k^2}+O(1/k^3)\right] 
\end{eqnarray*} 
where
\begin{eqnarray*} 
M_1(x,z)&=&-\frac{y(5y^4+(-12z+10)y^3+(12z^2+2-24z)y^2
+(-8z^2+12z-6)y+24z-4z^2-3)}{24(y+1)^2} 
\end{eqnarray*} 
and
\begin{eqnarray*} 
M_2(x,z)&=&\frac{y^2}{1152(y+1)^4 }
\left[385 y^8+(-840 z+1540) y^7+(1848+840 z^2-3360 z) y^6
+(-308+3040 z^2-4584 z-480 z^3) y^5
\right.
\nonumber\\
&+&\left.
(144 z^4-2306-1152 z^3+3304 z^2-1200 z) y^4+(-1408 z^2-192 z^4+3336 z-1524
+768 z^3) y^3
\right.
\nonumber\\
&+&\left. (24-32 z^4+1344 z^3+4128 z-4456 z^2) y^2+(-1632 z^2+324+2088 z
+64 z^4-288 z^3) y
\right.
\nonumber\\
&+&\left. 16 z^4+432 z-192 z^3+312 z^2+81
\right]
\end{eqnarray*} 
with
$$
y=\frac{1}{\sqrt{x^2+1}}.
$$
For the particular case $z=0$ we get
\begin{eqnarray*} 
M_1(x,0)&=&-\frac{y(5y^2-3)}{24} 
\end{eqnarray*} 
and
\begin{eqnarray*} 
M_2(x,0)&=&\frac{y^2}{1152}
\left[385 y^4-462y^2+81
\right]
\end{eqnarray*} 
which reprodices the known asymptotic formulae~\cite{abram}. 
This, as well as cancellation of
powers of $y$ in the denominators of $M_i$,
gives a quite reliable criterion for correctness
of the obtained expression.

\subsection{Asymptotic form of the Kummer function 
$M(k^2x^2/4z+1,k+2,z)$ for $k\gg 1$, $z, \ x$ fixed.}

Computations analogous to those of the previous section lead to the result
\begin{eqnarray*} 
M(k+1+a/4z,k+2,z)&=&
\frac{k!2^k}{(kx)^k}\frac{1}{\sqrt{2\pi k}}\frac{1}{(1+x^2)^{1/4}}\exp
\left\{  k\sqrt{x^2+1}+k\ln\left(\frac{x}{1+\sqrt{x^2+1}}\right)
\right\}  
\nonumber\\ 
&\times&\exp\left\{ \frac{z}{2}\frac{\sqrt{x^2+1}-1}{\sqrt{x^2+1}+1}\right\} 
\left[M_0(x)+\frac{M_1(x,z)}{k}+\frac{M_2(x,z)}{k^2}+O(1/k^3)\right] 
\end{eqnarray*} 
where
\begin{eqnarray*} 
M_0(x)&=&\frac{2y}{1+y},
\end{eqnarray*} 
\begin{eqnarray*} 
M_1(x,z)&=&-\frac{y(5 y^5+(-12 z+22) y^4+(-48 z+38+12 z^2) y^3
+(-36 z+6-8 z^2) y^2+(-39-4 z^2) y-24)}{12(y+1)^3} 
\end{eqnarray*} 
and
\begin{eqnarray*} 
M_2(x,z)&=&\frac{y^2}{576(y+1)^5 }
\left[385 y^9+(-840 z+2380) y^8+(-5040 z+6048+840 z^2) y^7
\right.
\nonumber\\
&+&\left. 
(-12744 z+7636-480 z^3+4480 z^2) y^6
\right.
\nonumber\\
&+&\left. (9160 z^2+144 z^4-16896 z-1728 z^3+3478) y^5+(-3660-192 z^4+6272 z^2
-10296 z) y^4
\right.
\nonumber\\
&+&\left. (-7440-32 z^4-1192 z^2+1728 z^3+1008 z) y^3+(-1728 z^2-5892+64 z^4
+480 z^3+4680 z) y^2
\right.
\nonumber\\
&+&\left.(1728 z+16 z^4-2439+408 z^2) y-432+192 z^2
\right]
\end{eqnarray*} 
with
$y=1/\sqrt{x^2+1}$.

\subsection{Asymptotic form of Kummer function $M(-k^2x^2/4z,k+1,-z)$ for 
$k\gg 1$, $z, \ x$ fixed.}
\label{Mkp2}

Use of previous calculations can be maximised by writing 
the function to be decomposed as
\begin{eqnarray*} 
M(-a/4z,k+1,-z)&=&e^{-z}M(a/4z+k+1,k+1,z)
\nonumber\\
&=&\frac{k!z^{-a/4z}2^{k+1-a/2z}}{\Gamma(a/4z)}
\frac{(4z)^{-k-1}\Gamma(a/4z)}{\Gamma(a/4z+k+1)}e^{-z}
\int_0^\infty ds s^{a/2z+k+1}e^{-s^2/4z}I_{k}(s).
\end{eqnarray*}
The final result is
\begin{eqnarray*} 
M(-a/4z,k+1,-z)&=&e^{-z}M(a/4z+k+1,k+1,z)=
\frac{k!2^k}{(kx)^k}\frac{1}{\sqrt{2\pi k}}\frac{1}{(1+x^2)^{1/4}}
\exp\left\{  k\sqrt{x^2+1}+k\ln\left(\frac{x}{1+\sqrt{x^2+1}}\right)
\right\}  
\nonumber\\ 
&\times&\exp\left\{ -\frac{z}{2}\frac{\sqrt{x^2+1}-1}{\sqrt{x^2+1}+1}\right\} 
\left[1+\frac{M_1(x,z)}{k}+\frac{M_2(x,z)}{k^2}+O(1/k^3)\right] 
\end{eqnarray*} 
where
\begin{eqnarray*} 
M_1(x,z)&=&-\frac{y(5y^4+(12z+10)y^3+(12z^2+2+24z)y^2+(-8z^2-12z-6)y
-24z-4z^2-3)}{24(y+1)^2} 
\end{eqnarray*} 
and
\begin{eqnarray*} 
M_2(x,z)&=&\frac{y^2}{1152(y+1)^4 }
\left[385 y^8+(840 z+1540) y^7+(840 z^2+1848+3360 z) y^6+(-308+3040 z^2+480 z^3
+4584 z) y^5
\right.
\nonumber\\
&+&\left.
(1152 z^3-2306+1200 z+144 z^4+3304 z^2) y^4+(-1408 z^2-192 z^4-3336 z-1524
-768 z^3) y^3
\right.
\nonumber\\
&+&\left. 
(24-4456 z^2-1344 z^3-32 z^4-4128 z) y^2+(-2088 z+324+288 z^3+64 z^4
-1632 z^2) y
\right.
\nonumber\\
&+&\left. 
16 z^4+192 z^3+81+312 z^2-432 z
\right]
\end{eqnarray*} 
with
$y=1/\sqrt{x^2+1}$.

\subsection{Asymptotic form of Kummer function $M(-k^2x^2/(4z),k+2,-z)$ 
for $k\gg 1$, $z$ fixed.}
The relation of $M(-k^2x^2/(4z),k+2,-z)$ to $M(-k^2x^2/(4z),k+1,-z)$ 
is analogous to that of $M(k^2x^2/(4z)+1,k+2,z)$ 
to $M(k^2x^2/(4z),k+1,z)$, 
thus the calculation is similar to that in Sec.\ref{Mkp2}.
The result is
\begin{eqnarray*} 
M(-a/4z,k+2,-z)&=& e^{-z}M(k+2+a/4z,k+2,z) \nonumber\\
&=&
\frac{k!2^k}{(kx)^k}\frac{1}{\sqrt{2\pi k}}\frac{1}{(1+x^2)^{1/4}}\exp\left\{  k\sqrt{x^2+1}+k\ln\left(\frac{x}{1+\sqrt{x^2+1}}\right)
\right\}  
\nonumber\\ 
&\times&\exp\left\{ -\frac{z}{2}\frac{\sqrt{x^2+1}-1}{\sqrt{x^2+1}+1}\right\} \left[M_0(x)+\frac{M_1(x,z)}{k}+\frac{M_2(x,z)}{k^2}+O(1/k^3)\right] 
\end{eqnarray*} 
where
\begin{eqnarray*} 
M_0(x)&=&\frac{2y}{1+y},
\end{eqnarray*} 
\begin{eqnarray*} 
M_1(x,z)&=&-\frac{y(5 y^5+(12 z+22) y^4+(48 z+12 z^2+38) y^3+(6-8 z^2-12 z) y^2+(-4 z^2-48 z-39) y-24)}{12(y+1)^3} 
\end{eqnarray*} 
and
\begin{eqnarray*} 
M_2(x,z)&=&\frac{y^2}{576(y+1)^5 }
\left[385 y^9+(2380+840 z) y^8+(5040 z+6048+840 z^2) y^7
\right.
\nonumber\\
&+&\left. 
(4480 z^2+7636+12264 z+480 z^3) y^6
\right.
\nonumber\\
&+&\left. 
(3478+1728 z^3+8008 z^2+144 z^4+13152 z) y^5+(-2376 z-1792 z^2-1152 z^3-192 z^4-3660) y^4
\right.
\nonumber\\
&+&\left. 
(-19056 z-32 z^4-11560 z^2-2112 z^3-7440) y^3+(-2880 z^2+672 z^3-13032 z-5892+64 z^4) y^2
\right.
\nonumber\\
&+&\left.
(384 z^3-2439+864 z+2712 z^2+16 z^4) y-432+192 z^2+2304 z
\right]
\end{eqnarray*} 
with
$$
y=\frac{1}{\sqrt{x^2+1}}.
$$

\end{document}